\documentclass[12pt]{article}

\usepackage{times,epsfig,graphics,amssymb,latexsym,amsmath,setspace,fullpage,algorithm,algorithmic,subcaption}
\usepackage{array,url}
\usepackage[usenames]{color}
\usepackage{enumitem}
\usepackage{multirow}
\usepackage[comma, numbers]{natbib}
\usepackage{bm} 
\usepackage{appendix,comment}
\usepackage{amsthm}
\usepackage{makecell}
\usepackage{abstract}

\newtheorem{theorem}{Theorem}[section]
\newtheorem{lemma}{Lemma}[section]

\newtheorem{proposition}{Proposition}[section]

\newtheorem{assumption}{Assumption}[section]

\newtheorem{remark}{Remark}[section]

\DeclareMathOperator*{\argmax}{arg\,max}

\newcommand{\G}{\mathcal{G}}
\newcommand{\A}{\mathbf{A}}

\newcommand{\CCC}{\mathcal{C}}

\newcommand{\R}{\mathbb R}
\newcommand{\XX}{\mathbf X}

\newcommand{\bbb}{{\boldsymbol \beta}}

\newcommand{\ppi}{\boldsymbol \pi}

\newcommand{\aaaa}{\mathbf a}

\newcommand{\DDDD}{\mathcal D}

\newcommand{\EEE}{\mathbb E}

\newcommand{\GGG}{\mathcal G}

\newcommand{\yy}{\mbox{$\mathbf y$}}

\newcommand{\PPP}{\mathcal P}

\newcommand{\s}{\mathbf s}

\newcommand{\SSSS}{\mathcal S}

\newcommand{\xx}{\mathbf x}

\newcommand{\rank}{\text{rank}}

\newcommand{\init}{\text{init}}

\newcommand{\curr}{\text{curr}}
\newcommand{\KL}{\text{KL}}

\newcommand{\KT}{\text{KT}}
\newcommand{\var}{\text{var}}
\newcommand{\obs}{\text{obs}}

\usepackage{todonotes, ulem}

\usepackage{booktabs}

\begin{document}
\title{\Large Model-free Rank Aggregation {in the Presence of Rater Heterogeneity}: A Maximum Score Approach}

\author{Haoran Zhang and  Yunxiao Chen\thanks{Haoran Zhang is with Department of Statistics and Data Science, Southern University of Science and Technology, Shenzhen, China
(e-mail: zhanghr@sustech.edu.cn).
Yunxiao Chen is with Department of Statistics, London School of Economics and Political Science, London, U.K. (e-mail:y.chen186@lse.ac.uk).
}}
\date{}

\maketitle

\begin{abstract}
This paper investigates the rank aggregation problem through the lens of multi-way comparison data derived from rater scores. Departing from traditional parametric frameworks—such as the Bradley-Terry and Plackett-Luce models—we propose a model-free method that accommodates highly heterogeneous preference distributions across raters and encompasses weak stochastic transitivity in pairwise comparisons as a special case.  
We establish the theoretical foundations of the proposed estimator by proving its consistency, demonstrating that the proportion of discordant pairs (Kendall’s tau) converges to zero in probability as the number of raters diverges. Furthermore, we derive upper and lower bounds for a performance metric based on Kendall's tau. In certain asymptotic regimes, these bounds coincide up to logarithmic factors, so the estimator is nearly minimax optimal. These results are obtained by analyzing the convergence behavior of a U-empirical process; the novel technical results developed for this analysis may be of independent theoretical interest. The practical utility of our method is validated through extensive simulations and applications to sports player rankings and survey preference aggregation.
\end{abstract}

{\small\noindent\textbf{Keywords:} Rater model, Kendall's tau,  minimax optimality, ranking, missing data, U-empirical process

\doublespacing

\section{Introduction} \label{Introduction}
Rank aggregation, the process of synthesizing multiple orderings or ratings into a single global ranking, is a fundamental problem in information retrieval and decision science. 
Models and methods for rank aggregation have broad applications in practice, ranging from ranking players or teams in sports to aggregating customers' preferences in marketing surveys. Traditionally, this task is typically solved using parametric models,  including the famous Thurstonian and Bradley-Terry (BT) models \cite{thurstone1927method, bradley1952rank} for pairwise comparison data, and the Plackett–Luce (PL) model \cite{luce1959individual,Plackett1975} for multi-way comparison data. Under parametric model assumptions, these models transform a rank aggregation problem, which is discrete in nature, into a parameter estimation problem that can be solved via continuous optimization. 
The statistical properties of the BT model have been established under both dense \cite{simons1999asymptotics} and sparse settings \cite{han2020asymptotic, gao2021uncertainty, han2023general}, as well as sequential settings where comparison pairs are adaptively chosen \cite{chen2022asymptotically}. More recently, theoretical results have also been established for the PL model \cite{fan2025ranking,Hanetal2025,fan2026spectral}.

In many real-world applications, the parametric assumptions of the Thurstonian, BT, and PL models can be overly simplified, resulting in biased rankings. Moving beyond restrictive parametric assumptions, nonparametric models and methods have been developed for pairwise comparison data, including rank aggregation based on Borda counting \cite{shah2018simple},  estimation of pairwise comparison probabilities under nonparametric BT models based on universal singular value thresholding \cite{chatterjee2015matrix} (USVT) and monotone constraints \cite{chatterjee2019estimation}, and noisy sorting \cite{braverman2008noisy,mao2018minimax}. To our knowledge, however, no nonparametric models or methods have been developed under general multi-way comparison settings. 

This paper proposes a MAximum Score esTimator for aggEgating Ranks (MASTER), a model-free approach for rank aggregation based on multi-way comparison data derived from raters’ scores. More specifically, we consider a system of $n$ items and $p$ raters, where each rater ranks a subset of the items, with the objective of recovering the latent global ranking. In the limiting case where each rater evaluates only two items, our framework reduces to the classical pairwise comparison setting.
The MASTER 
analyzes observed multi-way comparison data under a highly flexible setting, in which the data distribution is allowed to be highly heterogeneous across raters. This setting covers rank aggregation based on pairwise comparisons under the Weak Stochastic Transitivity (WST) assumption, as studied in \cite{mao2018minimax}.  
We establish the theoretical
foundations of the proposed estimator by proving its consistency, showing that the proportion of discordant pairs (Kendall’s tau)
converges to zero in probability as the number of raters diverges. 
Furthermore, we derive error bounds for a
performance metric based on Kendall’s tau, covering both dense and sparse regimes where the number of ratings per rater can be either a fixed constant or diverge with the total number of items. 
Under certain asymptotic regimes, these error bounds imply that the proposed estimator is nearly minimax optimal.
{Simulation studies in both pairwise and multi-way comparison settings show
that MASTER substantially improves over existing methods under heterogeneous
preference distributions and non-uniform sampling, while remaining competitive
when the BT or PL model is correctly specified. We further apply MASTER to a
professional tennis dataset based on pairwise match outcomes and a sushi
preference dataset based on multi-way partial rankings. The real-data analyses
show that MASTER yields interpretable aggregate rankings and reveals
meaningful departures from parametric rankings in the presence of preference
heterogeneity.}

The proposed method is closely related to the noisy sorting framework investigated by \cite{mao2018minimax}, yet there are fundamental differences. While \cite{mao2018minimax} addressed rank aggregation for pairwise comparisons under the WST and uniform-sampling assumptions, our work considers a significantly more complex setting involving multi-way comparisons and rater heterogeneity.
In our framework, the model space is considerably larger and more complex; notably, we allow a general sampling scheme, moving beyond the uniform-sampling requirements of prior work. Consequently, the theoretical machinery employed in \cite{mao2018minimax} to establish minimax optimality is insufficient for analyzing MASTER. In the special case of pairwise comparisons, our method coincides with an estimator used in \cite{mao2018minimax} for minimax analysis. By using more general proof techniques, our theoretical results not only recover the minimax rates established therein but also provide theoretical guarantees to broader, non-uniform, and multi-way comparison regimes.

In summary, the contributions of this paper are threefold. First, we establish a general framework for rank aggregation under significantly weaker assumptions and in more general settings than those in the existing literature. Our model-free approach encompasses traditional parametric models, such as the Bradley-Terry (BT) and Plackett-Luce (PL) models, as well as nonparametric BT variants based on Strong Stochastic Transitivity (SST). Unlike previous methods, such as noisy sorting, our framework does not rely on uniform sampling, offering greater flexibility in accounting for heterogeneity in rater behavior. Second, we provide rigorous error bounds that illuminate the information-theoretic fundamental limits and convergence rates of this rank aggregation problem, which, to the best of our knowledge, is the first model-free theory for rank aggregation from heterogeneous multi-way comparisons. The key technical contribution is a uniform concentration analysis for a nonstandard weighted U-empirical process indexed by permutations. The resulting estimator is nearly minimax optimal in the bounded-comparison-size regime and remains consistent under substantially more general sampling and preference heterogeneity than existing approaches.
Finally, we develop a practical estimation algorithm for the MASTER objective. Although maximizing the MASTER objective function is inherently NP-hard due to its combinatorial structure, we demonstrate through extensive simulations and real-data applications that our algorithm is empirically robust and yields high-quality rankings in practice.

The following notation is used in the rest. 
For an integer $n$, denote $[n] = \{1,...,n\}$ and define
$S_n$ as the permutation set of $[n]$. 
We write \(1\{\cdot\}\) for the indicator function
and \(|A|\) for the cardinality of a finite set \(A\).
Further define $S_n^+ = \{\ppi = (\pi_1,...,\pi_n): \pi_j =  1 + |\{i\in[n]: y_i<y_j\}|~~\text{for some  } \yy \in\R^n\}$ as all rank vector representations of weak orderings that allow ties.
For any vectors $\yy = (y_1,...,y_n)\in\R^n$, define the rank vector for $\yy$ as $\rank(\yy) = (\pi_1,...,\pi_n)\in S_n^+$, where $\pi_j = 1 + |\{i\in[n]: y_i<y_j\}|$ represents the rank of $y_j$ in an increasing order for $j\in[n]$. For example, for $\yy = (0.1, 0.2, 0.2, 0)$, we have $\rank(\yy) = (2, 3, 3, 1)$.
For two rankings
\(\ppi,\ppi'\in S_n\), define Kendall's tau distance as
\[
    \tau(\ppi,\ppi')
    =
    \left|
    \{(i,j):i<j,\;(\pi_i-\pi_j)(\pi'_i-\pi'_j)<0\}
    \right|.
\]
For positive sequences $\{a_n\}$ and $\{b_n\}$, we denote $a_n \lesssim b_n$ 
or, equivalently, $b_n \gtrsim a_n$,
if there exists a constant $c>0$ that $a_n \leq c b_n$ for all $n$. We also write $a_n \asymp b_n$ if $a_n \lesssim b_n$ and $a_n \gtrsim b_n$ hold. 

The remainder of the paper is organized as follows. Section~\ref{sec:method} introduces the MASTER framework and its underlying assumptions, followed by the derivation of upper and lower convergence bounds and an algorithm for its computation. In Section~\ref{sec:exp}, we evaluate the performance of the proposed method against several state-of-the-art competitors through extensive simulations and two real-world applications: ranking professional tennis players using pairwise comparisons and aggregating sushi preferences from multi-way comparison data. Finally, Section~\ref{sec:disc} concludes the paper with a summary of our contributions and a discussion of potential future directions.

\section{Proposed Method}\label{sec:method}

\subsection{Proposed MASTER method}\label{subsec:master}

Consider $n$ items ranked by $p$ raters, where each rater $k\in[p]$ ranks a subset of $n$ items, denoted as $\CCC_k\subseteq[n]$ with $L_k = |\CCC_k|$. Let $y_{ik}$ be the score on item $i$ given by rater $k$, allowed to be of different data types, such as continuous, ordinal, or count.  A higher score on an item than another indicates that the rater ranks it higher. 
Our goal is to derive a global ranking based on the observed data $y_{ik}$, $k=1, ..., p$, $i \in \mathcal C_k$. We define MASTER as follows:
\begin{equation}\label{eq:mse}
    \widehat\ppi = (\hat \pi_1, ..., \hat \pi_n)^\top \in \argmax_{\ppi\in S_n} L(\ppi),
\end{equation} 
where $\ppi = (\pi_1, ..., \pi_n) \in S_n$ is a permutation of $(1, ..., n)$
and $$L(\ppi)  = \sum_{k=1}^p \sum_{i<j\in \CCC_k} (1_{\{y_{ik} < y_{jk}\}}-1_{\{y_{jk}< y_{ik}\}}) 1_{\{\pi_i < \pi_j\}}$$
is our objective function. The output $\widehat \ppi$ gives the estimated global ranking for the $n$ items, where $\pi_i$ gives the ranking of item $i$ among the $n$ items. Specifically, $\hat \pi_i = n$ means item $i$ is ranked the highest among all the items, while $\hat \pi_i = 1$ means it is ranked the lowest. 
We provide a few remarks about this estimator.

\begin{remark}
Since the MASTER objective function depends exclusively on the relative rankings of items for each rater, observing the relative order of the items in $\mathcal C_{k}$ is sufficient to use MASTER. In other words, instead of observing the scores $y_{ik}$, we only need to observe the ranking of items in $\mathcal C_k$, i.e., a multi-way comparison, in order to use MASTER. 
In this setting, the scores $y_{ik}$ can be interpreted as the latent utilities of item $i$ for rater $k$, which in turn induce the observed ranking.
\end{remark}

\begin{remark}
When constructing the objective function, scores are compared only within the same rater, not across raters. 
In other words, the objective function only depends on the relative ranking of the items for each rater. 
This construction is key to accounting for the potentially high heterogeneity among raters, in that the same score may not mean the same thing to different raters. 
The rationale for MASTER is thus to identify a global ranking of the item that best matches the observed relative rankings among raters. 
\end{remark}

\begin{remark}

As the scores can be ordinal or count, it is possible that $y_{ik} = y_{jk}$, i.e., the rater ranks the two items in a tie.  This flexibility allows MASTER to handle ranking data with ties directly. In contrast, for certain parametric models, such as the BT and PL models, ties are not allowed.    

\end{remark}

\begin{remark}
    The objective function in \eqref{eq:mse} is closely related to the maximum score estimator for robust binary regression \cite{manski1975maximum, horowitz1992smoothed}, where we observe 
    independent observations $(y_i,\mathbf x_i)$ for $i=1,...,n$ with $y_i\in\{0,1\}$ and $\mathbf x_i\in \R^k$. The maximum score estimator takes the form as {$\hat{\boldsymbol\gamma} = \argmax_{\boldsymbol\gamma\in\R^k}\sum_{i=1}^n(2y_i - 1)1_{\{(\mathbf x_i^\top\boldsymbol\gamma)>0\}}$,} which takes a similar form as \eqref{eq:mse}. There are two differences between this maximum score estimator and the proposed MASTER. First, the summation in \eqref{eq:mse} is over pairs of items rather than over observations in the binary regression problem. Second, we focus on estimating the global ranking $\ppi$ rather than the regression coefficient vector $\boldsymbol\gamma$, and thus the sign of $\pi_i-\pi_j$ appears in the MASTER.
\end{remark}

\subsection{Theoretical properties of MASTER}

In what follows, we first introduce the assumptions underlying MASTER and then establish its theoretical properties.

\begin{assumption}[Extended WST (EWST)]\label{ass:mono}
    All $\{y_{ik}\}_{i\in\CCC_k,k\in[p]}$ are independent.
    There exists a global ranking $\ppi^* = (\pi_1^*,...,\pi_n^*)$ $\in S_n$ such that if $\pi_i^* < \pi_j^*$, then
\begin{equation*}\label{eq:mono}
\Pr(y_{ik}<y_{jk}) > \Pr(y_{jk}<y_{ik})\quad\text{for all}~~ k\in[p]. 
\end{equation*}
\end{assumption}

\begin{remark}[Connection with pairwise comparison under WST assumption {and rater homogeneity}]\label{rmk:wst} 
We consider the special case when $L_k = 2$, and each of the $p$ raters uniformly and independently selects and rates a pair from the $\binom{n}{2}$ possible pairs. Further, suppose raters are homogeneous in that $\Pr(y_{ik} < y_{jk}) = \Pr(y_{ik'} < y_{jk'})$ and $\Pr(y_{ik} > y_{jk}) = \Pr(y_{ik'} > y_{jk'})$ for all $i\neq j\in[n]$ and $k\neq k'\in[p]$.
Then, define a $n\times n$ matrix $W = (w_{ij})_{n\times n}$ where $w_{ij}$ is the number of raters that ranks item $i$ higher than item $j$, and $w_{ij} = 0$ if there does not exist a $k\in [p]$ such that $C_{k} = \{i,j\}$. 
Then the $W$ matrix can be seen as a pairwise comparison data matrix, and the success probabilities 
$p_{ij} = \Pr(y_{ik}>y_{jk})$ satisfy the WST assumption, a setting considered in \cite{mao2018minimax} for noisy sorting. 
Under this special case, 
the estimator \eqref{eq:mse} is equivalent to an estimator introduced in \cite{mao2018minimax} for establishing the minimax rate. 
However, as discussed in the sequel, the analysis of \eqref{eq:mse} is more involved because of within-rater dependence in our more general setting; that is, two terms associated with item pairs $(i,j)$ and $(i',j')$ in the summation of \eqref{eq:mse} are dependent if $i=i'$ or $j=j'$.
Given this connection with the WST assumption, Assumption~\ref{ass:mono} can be seen as an extension of the WST assumption to the current multi-rater and multi-way comparison setting, and thus is named an Extended WST (EWST) assumption. 
\end{remark}

\begin{remark}[Connection with nonparametric BT model]
We further compare the EWST assumption with the nonparametric BT model considered in \cite{chen2008extended,chatterjee2019estimation}. Consider the same pairwise comparison setting as in Remark \ref{rmk:wst}. The nonparametric BT model assumes the SST assumption in that if $\pi_i^* > \pi_j^* > \pi_l^*$, then $$p_{il} \geq \max\{p_{ij}, p_{jl}\} \geq 0.5.$$
In contrast, the WST only requires $p_{il}$, $p_{jl}$ and $p_{ij}$ all to be greater than or equal to 0.5.

\end{remark}

\begin{remark}[Connection with PL model]
It is also easy to see that the EWST assumption covers the PL model, which further reduces to the BT model under the pairwise comparison setting.  In fact, the PL model holds when  
 $y_{ik} = \beta_i + \epsilon_{ik}$, where $\beta_i$ is an item-specific parameter that does not depend on raters, and $\epsilon_{ik}$ for all $k\in[p]$ and $i \in \mathcal C_k$ are independent and identically distributed (i.i.d.), following a Gumbel distribution; see \cite{mcfadden1973conditional} and \cite{yellott1977relationship}. It is easy to see that the EWST assumption holds, with the latent global ranking given by $\text{rank}(\boldsymbol{\beta})$, for $\boldsymbol{\beta} = (\beta_1, ..., \beta_n)$. We note that the PL model does not account for rater heterogeneity, as it assumes $\epsilon_{ik}$s to be i.i.d. In contrast, no distributional assumption is made about $y_{ik}$s, which allows for highly heterogeneous preference distributions across raters. 
\end{remark}

Before presenting our main theoretical results, we provide some intuition about the proposed estimator. Consider the expectation of our objective function $l(\ppi)$, 
$$
\begin{aligned} 
\EEE L(\ppi) =& \sum_{k=1}^p\sum_{i<j\in\CCC_k}\left(\Pr(y_{ik}<y_{jk}) - \Pr(y_{jk}< y_{ik})\right)1_{\{\pi_i < \pi_j\}}.
\end{aligned}
$$
According to Assumption~\ref{ass:mono}, the multiplicative factor in each term of $\EEE L(\ppi)$, $\Pr(y_{ik}<y_{jk}) - \Pr(y_{jk}< y_{ik})$, is positive when $\pi_i^* < \pi_j^*$ and negative otherwise.
Thus, it is obvious that the expected objective function is maximized when $\ppi = \ppi^*$. 
In our theoretical analysis, we establish the concentration of 
$L(\ppi)$ to $\EEE L(\ppi)$, based on which we show the convergence of $\hat \pi$ towards $\pi^*$.

For $k\in[p]$, define $\PPP_k = \{(i,j):i<j\in\CCC_k\}$ as the set of item pairs compared by rater $k$.
For any $\ppi,\ppi'\in S_n$, define the discordant pair set between them as $$
B(\ppi,\ppi') = \{(i,j):~i<j ~\text{and}~
(\pi_i-\pi_j)(\pi_i'-\pi_j')<0\}.
$$
For $1\leq i<j\leq n$ and $k\in[p]$, define the probability gap, which quantifies the difficulty for rater $k$ to distinguish between items $i$ and $j$, as 
$
q_{ijk} = |\Pr(y_{ik}<y_{jk}) - \Pr(y_{jk}< y_{ik})|.
$
We assess the accuracy of $\hat \pi$ by a modified Kendall's tau distance 
$$
\tau_{m}(\widehat\ppi,\ppi^*) = \frac{(\sum_{k=1}^p\sum_{(i,j)\in B(\widehat\ppi,\ppi^*)\cap\PPP_k}q_{ijk})^2}{\sum_{k=1}^p|B(\widehat\ppi,\ppi^*)\cap \PPP_k|}.
$$
The metric $\tau_{m}$ is the intrinsic loss induced by the fixed sampling design and heterogeneous pairwise signal strengths. The Proposition II.1 below shows that, under random sampling, it controls the standard Kendall's tau distance up to the sampling intensity $\bar N$ and the minimum signal strength $q_{\min}$.




In Theorem~\ref{thm:upper} below, we establish an upper bound for the modified Kendall's tau metric. 

\begin{theorem}\label{thm:upper}
    Under Assumptions~\ref{ass:mono},
    $\tau_m(\widehat\ppi,\ppi^*) = O_p(L_{max}n\log n)$ as $p \to\infty$.
\end{theorem}

\begin{remark}
    The asymptotic property of $\tau_m(\widehat\ppi,\ppi^*)$ is related to a weighted U-empirical process 
    $
    G(\ppi) = \sum\limits_{k=1}^p \sum\limits_{i<j} a_{ijk} (z_{ijk} - \EEE z_{ijk}),
    $
    where $z_{ijk} = 1_{\{y_{ik} < y_{jk}\}}-1_{\{y_{jk}< y_{ik}\}}$ and $a_{ijk} = 
2\cdot 1_{\{(i,j)\in B(\ppi,\ppi^*)\cap\PPP_k\}}$. In the proof of Theorem~\ref{thm:upper}, we derive the stochastic orders for $\sup_{\ppi:\tau(\ppi,\ppi^*)=t}|G(\ppi)|$ for different $t$, and quantify the order that matches up with $L(\ppi^*) - L(\ppi)$.
When $L_k=2$, i.e. each rater only compares two items, $G(\ppi)$ reduces to a standard empirical process, as discussed in \cite{mao2018minimax}. 
However, when $L_k >2$, the technique used to bound $G(\ppi)$ is much more involved.
\end{remark}

\begin{remark}
Many existing works require uniform random sampling
design, such as \cite{mao2018minimax, chatterjee2015matrix}. 
In contrast, in establishing Theorem~\ref{thm:upper}, we treat $\CCC_k$ for $k\in[p]$ as fixed, which is much more flexible and allows non-uniform sampling.
Under the same setting of \cite{mao2018minimax} with $L_k = 2$ and each of the $p$ raters uniformly and independently selects a pair from the $\binom{n}{2}$ possible pairs, then Theorem~\ref{thm:upper} and the Proposition~\ref{prop:connect} below imply $\tau(\widehat\ppi,\ppi^*) = O_p(n^3\log n /(p q_{\min}^2))$, which matches up with the minimax rate established in \cite{mao2018minimax} up to a logarithm term.
\end{remark}


We relate the modified Kendall's tau distance to the standard Kendall's tau distance, under a flexible non-uniform random sampling design as specified by the following Assumption~\ref{ass:non_uniform}.

\begin{assumption}[Non-uniform random sampling design] \label{ass:non_uniform}
For each rater $k \in [p]$, let $\mathcal{S}_{n,k} = \{ \CCC \subseteq [n] : |\CCC| = L_k \}$ denote the set of all possible item subsets of size $L_k$. Rater $k$ independently selects a subset $\CCC_k \in \mathcal{S}_{n,k}$, and there exist universal constants $c_{\max} \ge c_{\min} > 0$ such that for any subset $\CCC \in \mathcal{S}_{n,k}$, the probability of selection satisfies:
\begin{equation*}
    c_{\min} \binom{n}{L_k}^{-1} \le \Pr(\CCC_k = \CCC) \le c_{\max} \binom{n}{L_k}^{-1}.
\end{equation*}
\end{assumption}

\begin{remark}
Assumption~\ref{ass:non_uniform} is a non-uniform random sampling design, which allows the probabilities for different subsets to be sampled to be different.
When $c_{\max} = c_{\min} = 1$, then it reduces to uniform sampling design in that the expected numbers of raters that rate different items are the same. 
When $L_k = 2$ for all $k\in[p]$, this further reduces to the uniform sampling scheme considered in \cite{mao2018minimax}.
Note that, for each rater $k$, Assumption~\ref{ass:non_uniform} requires the probabilities for different subsets $\CCC\in \SSSS_{n,k}$ to be sampled to be of the same order, which is a widely adopted assumption in the literature of pairwise comparison -\cite{simons1999asymptotics, chen2019spectral, han2020asymptotic}. 
\end{remark}

Define $N_{ij} = \sum_{k=1}^p 1_{\{\{i,j\}\subseteq \CCC_k\}}$ as the number of raters that rate both items $i$ and $j$, and $\bar{N} = \sum_{k=1}^p L_k(L_k-1)/(n(n-1))$ as the average number of total comparisons for different item pairs. Let $q_{\min} = \min_{k, i<j} q_{ijk} > 0$.
The following Proposition~\ref{prop:connect} establishes a non-asymptotic upper bound for the Kendall's tau distance under Assumption~\ref{ass:non_uniform}.

\begin{proposition}\label{prop:connect}
Suppose Assumptions~\ref{ass:mono} and \ref{ass:non_uniform} hold. Then, with probability at least $1 - \exp(-2n \log n)$, the standard Kendall's tau distance $\tau(\check\ppi, \ppi^*)$ for any data-dependent estimator $\check\ppi$ is bounded by:
\begin{equation*}
    \tau(\check\ppi, \ppi^*) \le \max \left\{ \frac{2 \tau_m(\check\ppi, \ppi^*)}{c_{\min} q_{\min}^2 \bar{N}}, \frac{14 L_{max}^2 n \log n}{c_{\min} \bar{N}} \right\}.
\end{equation*}
\end{proposition}

We now specify certain scenarios under which the proportion of discordant pairs converges to zero. 
\begin{proposition}\label{prop:KT}
    Suppose Assumption~\ref{ass:mono} holds. 
    Then, $\tau(\widehat\ppi,\ppi^*)/\binom{n}{2}=o_p(1)$, when $p\to\infty$ in the first scenario and $n\to\infty$ in the second scenario.
    \begin{enumerate}
        \item Assumption~\ref{ass:non_uniform} holds and $\bar Nq_{\min}^2 \gg L_{\max}\log n/n$.
        

        \item There is only one rater who ranks all $n$ items, e.g. $p=1$ and $L_1 = n$, and $\sum_{i,j\in[n]}\Pr(y_{i1} > y_{j1})1_{\{\pi_i^*<\pi_j^*\}}/\binom{n}{2} = o(1)$.
    \end{enumerate}
\end{proposition}

\begin{remark}
Under the random sampling design, a sufficient condition for the proportion of discordant pairs to converge to zero is $\bar Nq_{\min}^2 \gg L_{\max}\log n/n$, where the quantity $\bar Nq_{\min}^2$ could be seen as a measure of signal strength. 
    Under a special circumstance where there is only one rater who ranks all $n$ items, e.g. $p=1$ and $L_1 = n$, the rank consistency relies on the quantity $\sum_{i,j\in[n]}\Pr(y_{i1} > y_{j1})1_{\{\pi_i^*<\pi_j^*\}}/\binom{n}{2}$, which is the averaged inversion probabilities across all item pairs. If it decays to 0 when $n$ grows, then the proportion of discordant pairs converges to zero in probability as $n$ grows to infinity.
\end{remark}


As mentioned in Section~\ref{subsec:master}, the discrete optimization \eqref{eq:mse} contains $n$! elements in its feasible space, making it a challenge to find the global maximum. However, we point out that we do not have to find the global maximum; instead, an approximate solution $\widetilde\ppi$ satisfying $L(\widetilde\ppi)\geq L(\ppi^*)$ with high probability is sufficient. We formalize this in Proposition~\ref{prop:cond} below, the proof of which is given in the Appendix. 

\begin{proposition}\label{prop:cond}
    Under Assumption~\ref{ass:mono}, for any $\widetilde\ppi$ such that $L(\widetilde\ppi)\geq L(\ppi^*)$ with probability approaching 1, we have $\tau_m(\widetilde\ppi,\ppi^*) = O_p(L_{\max}n\log n)$.
\end{proposition}

In the following, we turn to establish the minimax lower bound for the rank estimator with respect to the modified Kendall's tau distance.
To this end, suppose that for each rater $k\in[p]$, we only observe the ranking of his ratings on items in $\CCC_k$, e.g. we observe $\DDDD = \{(\CCC_k,\ppi_{\obs,k})\}_{k=1}^p$, where $\ppi_{\obs,k} = \rank(\yy_{\CCC_k,k})\in S_{L_k}^+$.
Consider estimators based on $\DDDD$, e.g. $\check\ppi = T(\DDDD)$.
We establish the minimax lower bound under the random sampling design of Assumption~\ref{ass:non_uniform}.
Define the parameter space for the distribution of $\DDDD$ as $\GGG$ which consists of all joint distributions for $\DDDD$ satisfying Assumptions~\ref{ass:mono} and \ref{ass:non_uniform}.
For any $G\in \GGG$, due to the data generating process for $\DDDD$, there exists a unique ranking, denoted as $\ppi(G)\in S_n$, such that $\Pr(y_{ik}<y_{jk}) > \Pr(y_{jk}< y_{ik})$ for all $i,j\in \CCC_k$ and $k\in[p]$.
In the following, we consider to derive a lower bound for $\min_{\check\ppi = T(\DDDD) }\max_{G\in\GGG}\tau_{m}(\check\ppi,\ppi(G))$.

\begin{theorem}\label{thm:lower}
Fix any \(\delta\ge 1\) such that
the constants in Assumption~\ref{ass:non_uniform} satisfy \(\delta^{-1}\leq c_{\min}\le c_{\max}\le \delta\).
    If $\sum_{k=1}^pL_k(L_k-1)\leq n(n-1)/(40\delta\log(n+1))$, then
    \begin{equation}\label{eq:lower1}
        \min_{\check\ppi = T(\DDDD) }\max_{G\in\GGG}\EEE_{G}\tau_{m}(\check\ppi,\ppi) \gtrsim \frac{n}{\delta^2\log n}.
    \end{equation} 
    If $\sum_{k}L_k \leq n/\log (n+1)$, then $\min_{\check\ppi = T(\DDDD) }\max_{G\in\GGG}\EEE_{G}\tau_{m}(\check\ppi,\ppi) \gtrsim \sum_kL_k^2/\delta$.
\end{theorem}

\begin{remark}
We give a discussion for result of Theorem~\ref{thm:lower}. 
Under the regime $p\ll n^2/\log n$ and 
circumstance when $L_{\max} = O(1)$, which corresponds to pairwise comparison ($L_{\max} = 2$) and multiple comparison ($L_{\max}>2$), \eqref{eq:lower1} matches up with upper bound in Theorem~\ref{thm:upper} up to a logarithm factor,
which implies the nearly minimax optimality of the proposed method. 
Although Theorem~\ref{thm:lower} is stated in terms of the modified Kendall's tau
metric, its proof also shows that, when
\(\sum_{k=1}^p L_k\le n/\log(n+1)\),
\[
    \inf_{\check{\ppi}=T(D)}
    \sup_{G\in\mathcal G}
    \EEE_G\tau(\check{\ppi},\ppi(G))
    \asymp n^2.
\]
Thus, this is an information-theoretically impossible regime for rank aggregation in the usual Kendall's tau sense. The corresponding lower bound
under the modified Kendall's tau metric is
\[
    \inf_{\check{\ppi}=T(D)}
    \sup_{G\in\mathcal G}
    \EEE_G\tau_m(\check{\ppi},\ppi(G))
    \gtrsim
    \frac{\sum_{k=1}^p L_k^2}{\delta},
\]
which is the second claim in Theorem~\ref{thm:lower}.
\end{remark}




\subsection{Computation}

The objective function for MASTER in \eqref{eq:mse} involves an NP-hard discrete optimization problem. To find a good solution to \eqref{eq:mse}, we develop a greedy local search algorithm and combined it with an initial ranking obtained via smoothing. 
Our initialization 
is 
motivated by a smoothed maximum score estimator for binary regression \cite{horowitz1992smoothed}.
Let $w_{ij}$ and $l_{ij}$ be the numbers of raters that rank item $i$ higher and lower than item $j$, respectively. Then the objective function for MASTER can be written as $L(\ppi) = \sum_{i<j} (l_{ij} - w_{ij})1_{\{\pi_i<\pi_j\}}$. 
We consider the following smooth optimization
\begin{equation}\label{eq:surrogate}
\bbb_{\init} = \argmax_{\bbb\in\R^n} \sum_{i<j}(l_{ij} - w_{ij})\frac{e^{(\beta_j - \beta_i)/h}}{1 + e^{(\beta_j - \beta_i)/h}},\quad\text{subject to}\quad \sum_{i=1}^n\beta_i=0,
    \quad
    \frac{1}{n}\sum_{i=1}^n\beta_i^2=1.
\end{equation} 
where we replace the indicator function $1_{\{\pi_j-\pi_i > 0\}}$ in $L(\ppi)$ by the function $\phi_h(x)=(1+\exp(-x/h))^{-1}$ for some small $h>0$ to facilitate smooth optimization. 
{Note that as $h\to0$, $\phi_h(x)\to 1$ if $x>0$, and $\phi_h(x)\to 0$ if $x<0$. Therefore we use the objective function in \eqref{eq:surrogate} to approximate $L(\cdot)$.}
The constraints fix the scale of \(\bbb\), so that the smoothing parameter \(h\) cannot be absorbed by rescaling. We solve this problem sequentially for $h=1, 0.5, 0.25, 0.1,0.05,0.01$
using the solution from the previous value of \(h\) as the warm start for the next one. After each gradient ascent step, \(\bbb\) is re-centered and
re-scaled by
\[
    \bbb
    \leftarrow
    \frac{\sqrt n(\bbb-\bar\beta{\bf 1})}
    {\|\bbb-\bar\beta{\bf 1}\|_2},
    \qquad
    \bar\beta=\frac{1}{n}\sum_{i=1}^n\beta_i.
\]
Our initial ranking $\ppi_{\text{init}}$ is then obtained by $\ppi_{\init} = \text{rank}(\bbb_{\init})$. 
After getting $\ppi_{\init}$, we perform a consecutive $K$-tuple searching algorithm to obtain the final ranking, as described in Algorithm~\ref{algo} below.

\begin{algorithm}
\caption{$K$-tuple searching from an initial ranking based on surrogate objective} \label{algo}
\begin{tabbing}
{\bf Input:} \qquad \= numbers of winnings $(w_{ij})_{i<j}$, number of losses $(l_{ij})_{i<j}$, length of the tuple $K$\\
{\bf Initialize:} \qquad \> set $\ppi_{\curr}=\ppi_{\init}$ with $\ppi_{\init}$ from \eqref{eq:surrogate} and $t = K$ \\
{\bf While} $t\leq n$: \\
\> try all permutations of the rank segment $\{t-K+1,...,t\}$ \\
\> {\bf If} any improves objective $L(\cdot)$: \\
\> \qquad update $\ppi_{\curr}$ by applying the best permutation,  and set $t=K$ \\
\> {\bf Else} \\
\> \qquad update $t$ by $t+1$\\
{\bf Output:} $\ppi_{\curr}$
\end{tabbing} 
\end{algorithm}
The feasible space of this $n$-dimensional discrete optimization problem \eqref{eq:mse} consists of $n!$ elements. There is no guarantee that Algorithm~\ref{algo} finds the global optimal solution. However, as shown in Proposition~\ref{prop:cond}, we only need $\ppi_{\curr}$ to satisfy $L(\ppi_{\curr})>L(\ppi^*)$ with high probability. According to our simulation study with $ K=3$, the estimates given by Algorithm~\ref{algo} satisfy this condition in more than $95\%$ of experiments.

\section{Numerical Experiments}\label{sec:exp}

\subsection{Simulation study}\label{subsec:simu}


\textbf{Pairwise comparison.}
We first consider a pairwise comparison setting, which
corresponds to the special case \(L_k=2\) in our general framework. 
For each simulation setting, we generate $p=n^2$
raters. To allow non-uniform comparison frequencies, we first generate pair-specific sampling weights $\xi_{ij}\sim {\rm Uniform}(0.3,0.5)$ for $1\le i<j\le n$.
Then each rater independently selects one item pair according to
\[
    \Pr(\CCC_k=\{i,j\}\mid \{\xi_{ab}\})
    =
    \frac{\xi_{ij}}{\sum_{1\le a<b\le n}\xi_{ab}},
    \qquad k=1,\ldots,p.
\]
This design is a pairwise instance of Assumption~\ref{ass:non_uniform}, since the sampling
probability of each pair is of order \({n\choose 2}^{-1}\). Let $N_{ij}=\sum_{k=1}^p 1_{\{\CCC_k=\{i,j\}\}}$
denote the realized number of comparisons between items \(i\) and \(j\).
Given the selected pair \(\CCC_k=\{i,j\}\) with \(i<j\), we generate the comparison outcome
\[
    Z_k=1_{\{\text{rater }k\text{ ranks item }j\text{ above item }i\}}\quad
\text{from}\quad
    Z_k\sim {\rm Bernoulli}(p_{jik}),
\]
where the rater-specific comparison probabilities \(p_{jik}\) are generated
according to one of the following scenarios.
Throughout the three scenarios, we set the true global ranking as
\(\pi_i^*=i\).

\begin{enumerate}
\item Random heterogeneity. For each observed pair \(\CCC_k=\{i,j\}\) with \(i<j\), generate $p_{jik}\sim {\rm Uniform}(0.55,0.95)$.

\item Group-structured heterogeneity. Divide the items into two groups, \(\{1,\ldots,n/2\}\) and
\(\{n/2+1,\ldots,n\}\). For each observed pair \(C_k=\{i,j\}\) with
\(i<j\), generate
\[
    p_{jik}\sim
    \begin{cases}
    {\rm Uniform}(0.75,0.85), & \text{if } i,j \text{ are in the same group},\\
    {\rm Uniform}(0.65,0.75), & \text{otherwise}.
    \end{cases}
\]

\item BT model. Generate latent scores \(x_1,\ldots,x_n\) independently from \(N(0,2)\)
and sort them in increasing order as
\(x_{(1)}<\cdots<x_{(n)}\). Set
\[
    p_{jik}
    =
    \frac{1}{1+\exp[-(x_{(j)}-x_{(i)})]},
    \qquad i<j.
\]
\end{enumerate}

It is not difficult to see that all three scenarios satisfy the pairwise
version of the EWST assumption. In the first two scenarios, for every
observed pair \(C_k=\{i,j\}\) with \(i<j\), the rater-specific comparison
probability satisfies \(p_{jik}>1/2\), and hence item \(j\) is more likely
to be ranked above item \(i\) by rater \(k\). However, these two scenarios
do not satisfy SST in general. In the first scenario, this is due to the
randomness and heterogeneity in generating the rater-specific pairwise
comparison probabilities. In the second scenario, the violation is more
systematic: comparison probabilities tend to be closer to \(1/2\) when the
rank difference between two items is larger, i.e., when they are not in the
same group, which is opposite to the monotonicity required by SST. In
contrast, the third scenario follows a BT model. 
The comparison
probability \(p_{jik}\) is monotone increasing in \(x_{(j)}-x_{(i)}\).
Therefore, the third scenario satisfies SST and reduces to the standard
homogeneous BT model.

We compare the proposed MASTER with competitors designed for pairwise comparison data including the maximum likelihood estimator based on the BT model in \cite{han2020asymptotic}, the Borda counting algorithm proposed in \cite{shah2018simple}, and
universal singular value thresholding (USVT) \cite{chatterjee2015matrix}.
For the USVT method, we apply the algorithm in \cite{chatterjee2015matrix} to the standardized skew-symmetric matrix $\XX = (x_{ij})_{n\times n}$ with $$
x_{ij} = \left\{ 
\begin{aligned}
    &\frac{2w_{ij}}{w_{ij}+l_{ij}} - 1,\quad &&\text{if~~} w_{ij} + l_{ij} > 0,\\
    &0,\quad&& \text{if~~} w_{ij} + l_{ij} = 0,
\end{aligned}
\right.
$$
to estimate the pairwise comparison probabilities, and then rank the players by the row sums of the estimated pairwise comparison probability matrix. 

By combining the choices of item size with the three scenarios for comparison probabilities, we obtain nine simulation settings. For each setting, we generate 100 independent datasets. For each dataset, the methods mentioned previously are applied. For the MASTER, we run Algorithm~\ref{algo} with search size $K =3$. The accuracy of the aggregated ranking is measured by the standard Kendall's tau ranking error, which is calculated as the proportion of discordant pairs, i.e.,  $2\tau(\hat\pi,\pi^*) / (n(n-1))$. 
The results are summarized in Table~\ref{tab:pairwise_errors}.

\begin{table}[!htbp]
\centering
\caption{The averaged Kendall's tau ranking errors for different methods with standard errors based on 100 independent experiments. All error values are multiplied by 100 for presentation.} 
\label{tab:pairwise_errors}
\begin{tabular}{|c|c|c|c|c|c|c|c|}
\hline\hline
Scenarios & n & Borda  & BT & USVT  & MASTER
\\
\hline\hline
\multirow{3}{*}{Scenario 1} & 100 & 8.94 (0.06) & 6.37 (0.06) & 8.32 (0.08)   & \textbf{2.25 (0.05)}
\\
\cline{2-6}
 & 200 & 7.07 (0.03)  & 4.80 (0.03) & 6.30 (0.04) &   \textbf{1.25 (0.02)}
\\
\cline{2-6}
 & 500 & 4.90 (0.01)  & 3.10 (0.01) & 3.77 (0.02)   & \textbf{0.49 (0.00)}
\\
\hline\hline
\multirow{3}{*}{Scenario 2} & 100 & 12.52 (0.06)  & 8.65 (0.06) & 8.78 (0.08)   & \textbf{1.66 (0.05)}
\\
\cline{2-6}
 & 200 & 8.74 (0.03)  & 6.80 (0.04) & 7.79 (0.05)   & \textbf{0.85 (0.02)}
\\
\cline{2-6}
 & 500 & 6.70 (0.01)  & 5.39 (0.01) & 6.56 (0.02)  & \textbf{0.32 (0.00)}
\\
\hline\hline
\multirow{3}{*}{Scenario 3} & 100 & 7.10 (0.04)  & \textbf{3.76 (0.04)} & 4.67 (0.05)   & 6.48 (0.05)
\\
\cline{2-6}
 & 200 & 4.90 (0.02)   & \textbf{2.87 (0.02)} & 3.48 (0.02)  & 5.03 (0.03)
\\
\cline{2-6}
 & 500 & 3.08 (0.01)  & \textbf{1.75 (0.01)} & 2.17 (0.01)   & 3.02 (0.01)
\\
\hline\hline
\end{tabular}
\end{table}

Table~\ref{tab:pairwise_errors} shows that the proposed method substantially outperforms the competitors, achieving an error reduction by factors ranging from 2.8 to 16.8 compared to the second-best method. The advantage is more obvious in scenario 2, where the SST assumption is violated more severely. These results are due to the fact that all competitors assume SST. When the SST assumption is violated, the aggregated ranks from these methods are biased, resulting in higher errors compared to the proposed method. On the other hand, in scenario 3, where SST holds, all three competitors perform slightly better than the MASTER, and unsurprisingly, the BT model, under which the data are generated, performs the best. However, we should note that the MASTER is only slightly inferior to its competitors, due to the efficiency loss caused by having a larger parameter space. We should also note that the MASTER remains consistent in scenario 3, as SST implies WST.

\textbf{Multi-way comparison.}
We next consider a multi-way comparison setting. For
each simulation setting, we generate \(p=n^2\) raters. Let \(m=n/2\), and
divide the items into two groups \(G_1=\{1,\ldots,m\}\) and
\(G_2=\{m+1,\ldots,n\}\). We generate item-specific sampling weights
\(\xi_i\sim\text{Uniform}(0.5,1.5)\), \(i\in[n]\). The number of items
evaluated by rater \(k\) is generated independently from
\(L_k\sim\text{Uniform}\{3,4,5\}\). Conditional on \(L_k\), with probability
\(1-\eta\), where \(\eta=0.2\), we first draw
\(g_k\sim\text{Uniform}\{1,2\}\) and then sample \(\CCC_k\subset G_{g_k}\)
with \(|\CCC_k|=L_k\) according to
\[
    \Pr(\CCC_k=\CCC\mid g_k,L_k,\{\xi_i\}_{i=1}^n)
    =
    \frac{\prod_{i\in\CCC}\xi_i}
    {\sum_{\CCC'\subset G_{g_k}:|\CCC'|=L_k}\prod_{i\in\CCC'}\xi_i}.
\]
With probability \(\eta\), we sample \(\CCC_k\subset[n]\) with
\(|\CCC_k|=L_k\) according to
\[
    \Pr(\CCC_k=\CCC\mid L_k,\{\xi_i\}_{i=1}^n)
    =
    \frac{\prod_{i\in\CCC}\xi_i}
    {\sum_{\CCC'\subset[n]:|\CCC'|=L_k}\prod_{i\in\CCC'}\xi_i}.
\]
This design makes within-group comparisons more frequent while still
allowing cross-group comparisons.

Given \(C_k\), we generate latent scores according to one of the following
scenarios and observe only the induced ranking \(\ppi_{\text{obs},k}=\text{rank}(\yy_{\CCC_k,k})\). 
We
compare MASTER with the maximum likelihood estimator under the
Plackett--Luce model \cite{Hanetal2025} and a Borda counting method \cite{chen2022topk}. 
Throughout the three scenarios, the true global ranking is \(\pi_i^*=i\)
and \(\theta_i=3i/n\).

\begin{enumerate}
\item  Group-structured heteroscedastic scores. Given \(\CCC_k\), generate
\(y_{ik}=\theta_i+\sigma_{g(i)}\epsilon_{ik}\) for \(i\in\CCC_k\), where
\(\epsilon_{ik}\sim N(0,1)\) independently, \(g(i)\in\{1,2\}\) is the group
label of item \(i\), and \(\sigma_1=0.3,\sigma_2=3\).

   \item  Bottom-block heavy-tail score model. Let
\(B=\{1,\ldots,\lfloor bn\rfloor\}\) with \(b=0.25\). Given \(\CCC_k\),
generate
$$
y_{ik}=\theta_i+M1_{\{i\in B\}}\eta_{ik}+ J1_{\{i\notin B\}}+\sigma\epsilon_{ik},
$$
where \(\eta_{ik}\sim\text{Bernoulli}(q)\),
\(\epsilon_{ik}\sim N(0,1)\), independently across \(i\in\CCC_k\) and \(k\in[p]\).
We use \(q=0.3\), \(M=10\), $J = 0.05$ and \(\sigma=0.03\).

   \item  Plackett--Luce model. Given \(\CCC_k\), generate
\(y_{ik}=\theta_i+\epsilon_{ik}\) for \(i\in\CCC_k\), where
\(\epsilon_{ik}\sim\text{Gumbel}(0,1)\) independently across \(i\) and \(k\).
Then the induced ranking on \(\CCC_k\) follows a Plackett--Luce model with
item parameters \(\{\exp(\theta_i):i\in\CCC_k\}\).
\end{enumerate}

All three scenarios satisfy the EWST assumption. Scenario 1 introduces
group-dependent heteroscedasticity and therefore violates the homogeneous
PL structure. Scenario 2 further creates a top-choice bias: items in the
bottom block occasionally receive large shocks and appear near the top of
some observed rankings.
The additional shift \(J\) for items outside the bottom block ensures that the pairwise majority direction remains consistent with the true ranking \(\ppi^*\), so the EWST condition is still satisfied.
Scenario 3 is correctly
specified by the Plackett--Luce model and serves as a favorable benchmark
for the PL MLE.

\begin{table}[!htbp]
\centering
\caption{The averaged Kendall's tau ranking errors for multi-way comparison methods with standard errors based on 100 independent experiments. All error values are multiplied by 100 for presentation.} 
\label{tab:multiway_errors}
\begin{tabular}{|c|c|c|c|c|c|}
\hline\hline
Scenarios & n & Borda  & PL-MLE & MASTER
\\
\hline\hline
\multirow{3}{*}{Scenario 1} & 100 & 34.03 (0.02)   & 20.41 (0.07) & \textbf{8.90 (0.13)}
\\
\cline{2-5}
 & 200 & 36.56 (0.01)   & 19.54 (0.03) & \textbf{6.82 (0.06)}
\\
\cline{2-5}
 & 500 & 35.51 (0.00)  & 18.77 (0.02) & \textbf{4.80 (0.03)}
\\
\hline\hline
\multirow{3}{*}{Scenario 2} & 100 & 32.23 (0.02) & 4.35 (0.03) & \textbf{0.07 (0.00)}
\\
\cline{2-5}
 & 200 & 34.05 (0.01)   & 5.38 (0.01) & \textbf{0.17 (0.00)}
\\
\cline{2-5}
 & 500 & 32.99 (0.00)  & 4.84 (0.01) & \textbf{0.18 (0.00)}
\\
\hline\hline
\multirow{3}{*}{Scenario 3} & 100 & 35.74 (0.03)   & \textbf{2.37 (0.02)} & 5.81 (0.07)
\\
\cline{2-5}
 & 200 & 34.33 (0.01)  & \textbf{1.80 (0.01)} & 4.59 (0.04)
\\
\cline{2-5}
 & 500 & 36.27 (0.01)  & \textbf{1.20 (0.00)} & 3.29 (0.01)
\\
\hline\hline
\end{tabular}
\end{table}

Table~\ref{tab:multiway_errors} shows that the Borda method performs poorly in all three
scenarios. This is mainly due to the non-uniform sampling design: stronger
items are more likely to be compared with stronger items, and weaker items
are more likely to be compared with weaker items. As a result, Borda-type
scores mainly reflect local winning rates rather than global ranking
positions. In the first two scenarios, MASTER substantially outperforms the
PL-MLE, with error reduction factors ranging from about \(2.3\) to \(62.1\).
This is because the PL structure is violated by the heterogeneous score
distributions in Scenario 1 and by the bottom-block heavy-tail effect in
Scenario 2, while the EWST assumption still holds. In contrast, Scenario 3
is correctly specified by the PL model, and hence the PL-MLE performs the
best. Nevertheless, MASTER remains competitive and substantially improves
over the Borda-type methods, reflecting the efficiency loss of using a
model-free estimator compared with a correctly specified parametric model.

\subsection{Real data analysis}


\textbf{Tennis data (pairwise comparison).}
We apply the same methods considered in the simulation in Section~\ref{subsec:simu}  to a professional tennis match dataset, which contains the results of all men's matches organized by the Association of Tennis Professionals (ATP) from 2000 to 2018, collected from http://www.tennis-data.co.uk. The data covers Grand Slams, ATP Masters 1000, and other professional events. The same data has been considered in \cite{han2020asymptotic} and \cite{lee2025pairwise}. 
After removing players who lost all games, the set contains 875 players, with 7.13\% of player pairs having competed at least once. {Here, each match is treated as a pairwise comparison event with \(L_k=2\); thus, the proposed method is directly applicable even though the data do not involve human raters in the usual sense.}

Table~\ref{tab:top10_tennis} gives the top 10 players ranked by each method. The top 10 lists provided by the
counting and USVT methods do not align well with people's impression about the top male tennis players. For example,  the USVT method ranks Roddick A. third, ahead of, for example, Djokovic, N., and the counting algorithm ranks Ferrer D. fourth, ahead of, for example, Murray, A., which does not seem sensible. In fact, we do not expect the counting and USVT methods to work well for the current data, because these methods are based on the assumption that the missingness probabilities for the competing pairs are uniform. This assumption is unlikely to hold here as the likelihood of two professional tennis players meeting in a match depends on their ranking. The BT model and the MASTER do not need this assumption.

\begin{table}[htbp]
\centering 
\caption{Top 10 tennis players under different methods}\label{tab:top10_tennis}
\begin{tabular}{llll}
\hline
Counting & USVT & BT & MASTER \\
\hline
Federer R. & Federer R. & Djokovic N. & Djokovic N. \\
Nadal R. & Nadal R. & Federer R. & Nadal R. \\
Djokovic N. & Roddick A. & Nadal R. & Federer R. \\
Ferrer D. & Djokovic N. & Murray A. & Murray A. \\
Murray A. & Murray A. & Roddick A. & Hewitt L. \\
Berdych T. & Hewitt L. & Agassi A. & Agassi A. \\
Roddick A. & Ferrer D. & del Potro J.M. & del Potro J.M. \\
Robredo T. & Berdych T. & Schuettler P. & Kuerten G. \\
Hewitt L. & del Potro J.M. & Tsonga J.W. & Zverev A. \\
Gasquet R. & Gasquet R. & Ancic I. & Wawrinka S. \\
\hline
\end{tabular}
\end{table}

The rankings produced by the MASTER and the BT model are
positively associated with substantial differences. Kendall's tau and  Spearman's rho rank correlations between the two rankings are 0.47 and 0.65, respectively. We provide some discussions about the differences between the top 10 player lists from the two methods. First, the rivalry between Nadal, R. and Federer, R. is one of the most significant in tennis history, which also appears here. The BT model ranks Federer as the second and Nadal as the third, while the MASTER reverses this ordering. Considering that Nadal has won 24 out of 40 head-to-head competitions against Federer during their careers, and 10 out of 14 Grand Slam competitions, we believe that ranking Nadal ahead of Federer makes more sense when SST is not assumed. Second,  
it is interesting to see that the BT model ranks Roddick, A., as the fifth and Hewitt, L., as the eleventh, while the MASTER reverses this ordering. We believe that Roddick's high ranking under the BT model is boosted by his strong head-to-head record against Djokovic, N., who is ranked first by both methods. 
Roddick and Djokovic competed nine times, out of which Roddick won five times. This strong record carries significant weight in the likelihood of the BT model, resulting in a high latent score estimate for Roddick. 
In contrast, Hewitt competed against Djokovic seven times, winning only once. To better understand the rankings, we check some head-to-head records. The record of Roddick against Hewitt is tied at 7:7. However, the records of Roddick and Hewitt against Agassi, A. and del Potro, J.M., who rank sixth and seventh by both methods, show that having Hewitt in the fifth place makes more sense. Specifically, the records are 1:5, 1:4, 4:4, and 3:2, for Roddick against Agassi, Roddick against del Potro, Hewitt against Agassi, and Hewitt against del Potro, respectively, which suggests that ranking Roddick ahead of Agassi and del Potro is not sensible. 

\textbf{Sushi data (multi-way comparison).}
We further apply MASTER to the Sushi Preference Dataset\footnote{Data are available through the link: \url{https://www.kamishima.net/sushi/}}, which contains the preference records of 5,000 users regarding a set of 100 different types of sushi, ranging from common seafood to regional specialties.
The data were collected through a large-scale questionnaire survey, in which each of the 5,000 respondents ranked 10 types of sushi from most to least preferred.
We compare MASTER with the estimate from the PL model.  


The Kendall's tau correlation between the estimated rankings from the MASTER and PL models is 0.81, indicating that the two methods exhibit a strong overall consensus on the dataset's broader preference structure. We further compare the top ten items given by the two methods, as shown in Table \ref{tab:top10_sushi}, where the sushi types are ordered from the first to the tenth for each method. 
As a reference, we also show the pairwise winning rates between the top ten sushi under MASTER in Table~\ref{tab:rates}, where each entry shows the proportion that the sushi type listed in the corresponding row is preferred over that in the corresponding column, among raters who have rated both types of sushi. For example,  it shows that ``toro" is preferred over ``chu\_toro" 49\% of the raters who have rated both. As shown in Table \ref{tab:top10_sushi}, 
the top two rankings for ``chu\_toro" and ``toro" are consistent for both methods. In addition, the ranking of the sushi types in the third to the tenth are also largely consistent. However, there are also notable differences. For example, the sushi type ``maguro" is ranked third by the PL model but ranked eighth by the MASTER. Its third rank by the PL model may be boosted by the PL model's parametric assumption. As we can see from 
Table~\ref{tab:rates}, the eighth ranking of ``maguro" perfectly aligns with the WST assumption, in the sense that its winning rate is below 0.5 when compared with the first seven sushi types and above 0.5 when compared with the last two. Another notable difference is the ranking of ``kurumaebi”, which 
is ranked fifth by the PL model but to the tenth by MASTER. 
The higher ranking under the PL model may be boosted by its high winning rate (0.69) against ``negi\_roro",  which is ranked high by both methods, under the restricted parametric assumption of the PL model. On the other hand, it is consistently less preferred over the types ``tarabagani", ``maguro", and ``toro\_samon" in the seventh to the ninth ranks under the MASTER, which leads to its tenth aggregated rank under the MASTER.

\begin{table}[htbp]
\centering
\caption{Top 10 sushi under different methods} \label{tab:top10_sushi}
\resizebox{\textwidth}{!}{ 
\begin{tabular}{l|llllllllll}
\hline
PL model & chu\_toro & toro & maguro & negi\_toro & amaebi & kurumaebi & negi\_toro\_maki & samon & tarabagani & tai \\
\hline
MASTER   & chu\_toro & toro & negi\_toro & negi\_toro\_maki & amaebi & samon & tarabagani & maguro & toro\_samon & kurumaebi \\
\hline
\end{tabular}
}
\end{table}

\begin{table}[htbp]
\centering
\caption{Winning rates between the top 10 sushi under MASTER} 
\label{tab:rates}
\resizebox{\textwidth}{!}{ 
\begin{tabular}{lcccccccccc}
  \hline
 & chu\_toro & toro & negi\_toro & negi\_toro\_maki & amaebi & samon & tarabagani & maguro & toro\_samon & kurumaebi \\ 
  \hline
chu\_toro &  & 0.51 & 0.63 & 0.75 & 0.64 & 0.62 & 0.68 & 0.65 & 0.70 & 0.69 \\ 
  toro & 0.49 &  & 0.61 & 0.69 & 0.66 & 0.60 & 0.69 & 0.68 & 0.72 & 0.62 \\ 
  negi\_toro & 0.37 & 0.39 &  & 0.56 & 0.57 & 0.57 & 0.49 & 0.51 & 0.79 & 0.31 \\ 
  negi\_toro\_maki & 0.25 & 0.31 & 0.44 &  & 0.52 & 0.51 & 0.50 & 0.57 & 0.62 & 0.20 \\ 
  amaebi & 0.36 & 0.34 & 0.43 & 0.48 &  & 0.57 & 0.48 & 0.51 & 0.55 & 0.62 \\ 
  samon & 0.38 & 0.40 & 0.43 & 0.49 & 0.43 &  & 0.58 & 0.53 & 0.49 & 0.48 \\ 
  tarabagani & 0.32 & 0.31 & 0.51 & 0.50 & 0.52 & 0.42 &  & 0.51 & 0.55 & 0.56 \\ 
  maguro & 0.35 & 0.32 & 0.49 & 0.43 & 0.49 & 0.47 & 0.49 &  & 0.57 & 0.55 \\ 
  toro\_samon & 0.30 & 0.28 & 0.21 & 0.38 & 0.45 & 0.51 & 0.45 & 0.43 &  & 0.57 \\ 
  kurumaebi & 0.31 & 0.38 & 0.69 & 0.80 & 0.38 & 0.52 & 0.44 & 0.45 & 0.43 &  \\ 
   \hline
\end{tabular}
}
\end{table}

\section{Discussions}\label{sec:disc}

In this paper, we propose a novel framework for rank aggregation that operates under substantially weaker assumptions than existing methodologies. We establish the theoretical properties of the proposed estimator, including its statistical consistency and error bounds. To facilitate implementation, we develop an efficient algorithm for computing the estimator and demonstrate its advantages over state-of-the-art methods through extensive simulation studies. Furthermore, the practical utility of our approach is validated by applying it to rank professional tennis players using data from the Association of Tennis Professionals (ATP) and to analyze sushi preference data. In both cases, the proposed method yields meaningful results. 

One limitation of the MASTER framework is the computational challenge posed by its discrete optimization objective. While the proposed search algorithm performs effectively within our simulation settings, its performance in higher-dimensional or more general configurations warrants further scrutiny. To address this, future work should explore more sophisticated stochastic search techniques, such as simulated annealing. By incorporating controlled randomness, such algorithms could better navigate the complex discrete landscape to escape local optima and more reliably identify the global maximizer.

A distinct advantage of our approach, which remains underexploited, is its natural compatibility with privacy-preserving randomization. Protecting rater privacy is paramount when handling preference data, as individual rankings often reveal sensitive personal information. As our assumptions are highly nonparametric, 
they are likely to hold even after data perturbation. Consequently, the MASTER approach may maintain its theoretical consistency and error bounds under privatization without the need for the complex debiasing procedures required by parametric models (see, e.g., \cite{xu2025rate}), whereas randomization often leads to violation of the model's structural assumptions. We leave the formal development of these privacy-preserving extensions for future investigation.

\bigskip \bigskip

\appendices

\section{Proof of Theoretical Results}

\begin{proof}[Proof of Proposition~\ref{prop:connect}]
For any fixed ranking $\ppi \in S_n$, define the random variable $X(\ppi) = \sum_{k=1}^p |B(\ppi, \ppi^*) \cap \mathcal{P}_k|$ representing the total number of observed discordant pairs for permutation $\ppi$. Let $W_k(\ppi) = |B(\ppi, \ppi^*) \cap \mathcal{P}_k|$ such that $X(\ppi) = \sum_{k=1}^p W_k(\ppi)$.

Under Assumption \ref{ass:non_uniform}, the marginal probability $p_{ij,k}$ that a specific pair $(i,j)$ is evaluated by rater $k$ is the sum of probabilities of all subsets of size $L_k$ containing both $i$ and $j$. Since there are $\binom{n-2}{L_k-2}$ such subsets, we have:
\begin{equation*}
    c_{\min} \frac{\binom{n-2}{L_k-2}}{\binom{n}{L_k}} \le p_{ij,k} \le c_{\max} \frac{\binom{n-2}{L_k-2}}{\binom{n}{L_k}},
\end{equation*}
which simplifies to $c_{\min} \frac{L_k(L_k-1)}{n(n-1)} \le p_{ij,k} \le c_{\max} \frac{L_k(L_k-1)}{n(n-1)}$. 

Let $\mu(\ppi) = \mathbb{E}[X(\ppi)]$ be the expectation. By the linearity of expectation, summing over the $\tau(\ppi, \ppi^*) = |B(\ppi, \ppi^*)|$ discordant pairs yields:
\begin{equation*}
    c_{\min} \tau(\ppi, \ppi^*) \bar{N} \le \mu(\ppi) \le c_{\max} \tau(\ppi, \ppi^*) \bar{N}.
\end{equation*}

Since each rater samples independently, $X(\ppi)$ is the sum of independent bounded random variables $W_k(\ppi) \in [0, \frac{L_{max}^2}{2}]$. Using the uncentered second moment to bound the variance, we have $\sum_{k=1}^p var(W_k(\ppi)) \le \frac{L_{max}^2}{2} \mu(\ppi)$. Applying Bernstein's inequality to the lower tail with $\epsilon = 1/2$, we obtain:
\begin{align*}
    \Pr\left(X(\ppi) \le \frac{1}{2} \mu(\ppi) \right) &\le \exp\left( \frac{- (1/2)^2 \mu(\ppi)^2}{2 \left(\frac{L_{max}^2}{2}\right) \mu(\ppi) + \frac{2}{3} \left(\frac{L_{max}^2}{2}\right) (1/2) \mu(\ppi)} \right) \\
    &= \exp\left( - \frac{3 \mu(\ppi)}{14 L_{max}^2} \right) \\
    &\le \exp\left( - \frac{3 c_{\min} \tau(\ppi, \ppi^*) \bar{N}}{14 L_{max}^2} \right).
\end{align*}

To address the data-dependency of the estimator $\check\ppi$, we establish a uniform bound over the permutation space. Define a threshold $t^* = \frac{14 L_{max}^2 n \log n}{c_{\min} \bar{N}}$ and consider the subset of permutations with large errors, $S_{\text{bad}} = \{\ppi \in S_n : \tau(\ppi, \ppi^*) \ge t^*\}$. 
Taking the union bound over all permutations in $S_{\text{bad}}$:
\begin{align*}
    \Pr\left(\exists \ppi \in S_{\text{bad}} \text{ s.t. } X(\ppi) \le \frac{1}{2} \mu(\ppi) \right) &\le \sum_{\ppi \in S_{\text{bad}}} \exp\left( - \frac{3 c_{\min} \tau(\ppi, \ppi^*) \bar{N}}{14 L_{max}^2} \right) \\
    &\le |S_n| \exp\left( - \frac{3 c_{\min} t^* \bar{N}}{14 L_{max}^2} \right).
\end{align*}
Since $|S_n| = n! \le \exp(n \log n)$ and $t^* = \frac{14 L_{max}^2 n \log n}{c_{\min} \bar{N}}$, the exponent simplifies to $-3 n \log n$. Thus, the probability is bounded by $\exp(n \log n) \exp(-3 n \log n) = \exp(-2n \log n)$.

Therefore, with probability at least $1 - \exp(-2n \log n)$, the following uniform condition holds for \textit{all} $\ppi \in S_n$:
Either $\tau(\ppi, \ppi^*) \le t^*$ or $X(\ppi) > \frac{1}{2} \mu(\ppi) \ge \frac{1}{2} c_{\min} \tau(\ppi, \ppi^*) \bar{N}$.

Recall the definition of the modified Kendall's tau distance:
\begin{equation*}
    \tau_m(\ppi, \ppi^*) = \frac{\left(\sum_{k=1}^p \sum_{(i,j) \in B(\ppi, \ppi^*) \cap \mathcal{P}_k} q_{ijk}\right)^2}{X(\ppi)} \ge \frac{q_{\min}^2 X(\ppi)^2}{X(\ppi)} = q_{\min}^2 X(\ppi).
\end{equation*}

If a permutation falls into the second case ($X(\ppi) > \frac{1}{2} c_{\min} \tau(\ppi, \ppi^*) \bar{N}$), substituting this into the metric definition yields:
\begin{equation*}
    \tau_m(\ppi, \ppi^*) > q_{\min}^2 \left( \frac{1}{2} c_{\min} \tau(\ppi, \ppi^*) \bar{N} \right) \implies \tau(\ppi, \ppi^*) < \frac{2 \tau_m(\ppi, \ppi^*)}{c_{\min} q_{\min}^2 \bar{N}}.
\end{equation*}

Because the data-dependent estimator $\check\ppi$ is a realization within $S_n$, it must satisfy at least one of these two conditions. Combining them yields the final bound:
\begin{equation*}
    \tau(\check\ppi, \ppi^*) \le \max \left\{ \frac{14 L_{max}^2 n \log n}{c_{\min} \bar{N}}, \frac{2 \tau_m(\check\ppi, \ppi^*)}{c_{\min} q_{\min}^2 \bar{N}} \right\}.
\end{equation*}
This completes the proof. 
\end{proof}

\begin{proof}[Proof of Theorem~\ref{thm:upper}]
Without loss of generality, assume $\pi_i^* = i$.
We have
$$
L(\ppi^*) - L(\ppi) = 2\sum_{k=1}^p \sum_{(i,j)\in B(\ppi,\ppi^*)\cap\PPP_k}(1_{\{y_{ik} < y_{jk}\}}-1_{\{y_{jk}< y_{ik}\}}),\quad \text{and}
$$
\begin{equation}\label{eq:signal}
\begin{aligned}
\EEE L(\ppi^*) - \EEE L(\ppi) =& 2\sum_{k=1}^p\sum_{(i,j)\in B(\ppi,\ppi^*)\cap\PPP_k}\left[\Pr(y_{ik}<y_{jk}) - \Pr(y_{jk}> y_{jk})\right]\\ 
=& 2\sum_{k=1}^p\sum_{(i,j)\in B(\ppi,\ppi^*)\cap\PPP_k}q_{ijk}\geq 0.
\end{aligned}
\end{equation}

Define $G(\ppi) = L(\ppi^*) - L(\ppi) - (\EEE L(\ppi^*) - \EEE L(\ppi))$ for $\ppi\in S_n$. 
In the following, for a given integer $t$, we give an upper bound for $\sup_{\ppi:\sum_{k=1}^p|B(\ppi,\ppi^*)\cap\PPP_k|=t}|G(\ppi)|$. 
For $i,j\in[n]$ and $k\in[p]$, define $z_{ijk} = 1_{\{y_{ik} < y_{jk}\}}-1_{\{y_{jk}< y_{ik}\}}$. 
For $k\in[p]$, let $\A_k = (a_{ijk})_{n\times n}$ where $a_{ijk} = 
2\cdot 1_{\{(i,j)\in B(\ppi,\ppi^*)\cap\PPP_k\}}$ and $a_{jik} = -a_{ijk}$ for $1\leq i<j\leq n$.
We start by noting that $$
G(\ppi) = \sum_{k=1}^p \sum_{i<j} a_{ijk} (z_{ijk} - \EEE z_{ijk}) =: \sum_{k=1}^p I_k.
$$ We bound each $I_k = \sum_{i<j} a_{ijk} (z_{ijk} - \EEE z_{ijk})$ separately. 
We need the results of the following lemma, whose proof is deferred.
\begin{lemma} \label{lem:tech}
We have
\begin{equation*}
\begin{aligned}
&\EEE (I_k \mid y_{ik}) = \sum_{j:j\neq i} a_{ijk}(\EEE(z_{ijk}\mid y_{ik}) - \EEE z_{ijk}), \\
&\sum_{i\neq j}a_{ijk}\EEE(z_{ijk}\mid y_{ik}) = \sum_{i\neq j} a_{ijk}\EEE (z_{ijk}\mid y_{jk}),\\
&I_k = \sum_{i<j}a_{ijk}(z_{ijk} - \EEE z_{ijk}) = \frac{1}{2}\sum_{i\neq j} a_{ijk} (z_{ijk} - \EEE z_{ijk}).
\end{aligned}
\end{equation*}
\end{lemma}

According to Lemma~\ref{lem:tech}, we have 
$$
\begin{aligned}
    I_k - \sum_{i=1}^n \EEE (I_k\mid y_{ik}) =& \frac{1}{2}\sum_{i\neq j} a_{ijk} (z_{ijk}-\EEE z_{ijk}) - \sum_{i\neq j} a_{ijk}(\EEE(z_{ijk}\mid y_{ik}) - \EEE z_{ijk}) \\
    =& \frac{1}{2}\sum_{i\neq j}a_{i j k}(z_{ijk} - \EEE(z_{ijk}\mid y_{ik}) - \EEE(z_{ijk}\mid y_{jk})+ \EEE z_{ijk}),
\end{aligned}
$$ 
which implies $I_k = I_{k1} + I_{k2}$ with $$
\begin{aligned}
&I_{k1} = \sum_{i=1}^n \EEE (I_k\mid y_{ik}),\\
&I_{k2} = \frac{1}{2}\sum_{i\neq j}a_{ijk}(z_{ijk} - \EEE(z_{ijk}\mid y_{ik}) - \EEE(z_{ijk}\mid y_{jk})+ \EEE z_{ijk}) =: \sum_{i\neq j} a_{ijk}h_2(y_{ik}, y_{jk}),
\end{aligned}
$$
where $h_2(y_{ik}, y_{jk}) = z_{ijk} - \EEE(z_{ijk}\mid y_{ik}) - \EEE(z_{ijk}\mid y_{jk})+ \EEE z_{ijk}$.
We then derive non-asymptotic bounds for both $\sum_{k=1}^pI_{k_1}$ and $\sum_{k=1}^pI_{k_2}$.

\textbf{Non-asymptotic bound for $\bm{\sum_{k=1}^pI_{k_1}}$.}
Note that $$
\begin{aligned}
\EEE z_{ijk} =& \Pr(y_{ik}<y_{jk}) - \Pr(y_{jk}> y_{jk}),\\
\EEE(z_{ijk}\mid y_{ik}) =& 2\EEE\left[1_{\{y_{ik}<y_{jk}\}}\mid y_{ik}\right] - \EEE\left[1_{\{y_{ik}\neq y_{jk}\}} \mid y_{ik}\right] \\
=& 2S_{jk}(y_{ik}) - H_{jk}(y_{ik}), 
\end{aligned}
$$ where $S_{jk}(y_{ik}) = 1 - F_{jk}(y_{ik})$ and $H_{jk}(y_{ik}) = \Pr(y_{ik}\neq y_{jk} \mid y_{ik})$. 
Then, 
$$
\begin{aligned}
\EEE(I_k\mid y_{ik}) =& \sum_{j:j\neq i}a_{ijk}\Big( 2[S_{jk}(y_{ik}) - \Pr(y_{ik}<y_{jk})] - [H_{jk}(y_{ik}) - \Pr(y_{ik}\neq y_{jk})] \Big)\\
=&\sum_{j:j\neq i}a_{ijk}(w_{ijk}-\EEE w_{ijk})\quad \text{with}~~ w_{ijk} :=2S_{jk}(y_{ik}) - H_{jk}(y_{ik}),
\end{aligned}
$$ such that $$
\begin{aligned}
|\EEE(I_k\mid y_{ik})| \leq& C\sum_{j}|a_{ijk}|\leq CL_k\leq CL_{\max}\\
\var(\EEE(I_k\mid y_{ik})) \leq& L_k\sum_{j:j\neq i} a_{ijk}^2\var(w_{ijk}) \leq CL_k\sum_{j:j\neq i}a_{ijk}^2. 
\end{aligned}
$$ 
Applying Bernstein's inequality to $\sum_{k=1}^pI_{k1} =\sum_{k=1}^p \sum_{i=1}^n \EEE(I_k\mid y_{ik})$, which is the summation of $np$ independent bounded random variables, yields,
\begin{equation}\label{eq:1st order}
\begin{aligned}
\Pr(|\sum_{k=1}^pI_{k1}|\geq \epsilon) \leq& 2\exp\left\{\frac{-\epsilon^2}{2\sum_{k=1}^p\sum_{i=1}^n\var(\EEE (I_k\mid y_{ik})) + 2CL_{\max}\epsilon/3 } \right\}\\
\lesssim& \exp\left\{ \frac{-\epsilon^2}{L_{\max}\sum_{k=1}^p\sum_{i\neq j}a_{ijk}^2 + L_{\max}\epsilon} \right\}  \\
\lesssim& \exp\left\{ \frac{-\epsilon^2}{ L_{\max}\sum_{k=1}^p|B(\ppi,\ppi^*)\cap\PPP_k| + L_{\max}\epsilon  }  \right\},
\end{aligned}
\end{equation}
where the last inequality is due to that $\sum_{i\neq j}a_{ijk}^2 = 8\sum_{i< j}1_{\{(i,j)\in B(\ppi,\ppi^*)\cap\PPP_k\}} = 8|B(\ppi,\ppi^*)\cap\PPP_k|$.


\textbf{Non-asymptotic bound for $\bm{\sum_{k=1}^pI_{k_2}}$.}
We now consider $I_{k2} = \sum_{i\neq j} a_{ijk}h_2(y_{ik}, y_{jk})$. Since $h_2(y_{ik},y_{ik}) = 0$, then $I_{k2} = \sum_{i,j} a_{ijk}h_2(y_{ik}, y_{jk})$. 
By decoupling theorem (e.g. Theorem 3.1.2 in \cite{de2012decoupling}), we have $$
\EEE_{y}\exp(\lambda\sum_{i,j}a_{ijk}h_2(y_{ik},y_{jk}))\leq C\EEE_{y,y'}\exp(\lambda\sum_{i,j}a_{ijk}h_2(y_{ik},y_{jk}'))\quad\text{for}~\lambda > 0,
$$ where $y_{ik}'$ are independent copies of $y_{ik}$.
We first consider $\EEE_{y'}\exp(\lambda\sum_{i,j}a_{ijk}h_2(y_{ik},y_{jk}'))$. We have $$
\begin{aligned}
    &\EEE_{y'}\exp(\lambda\sum_{i,j}a_{ijk}h_2(y_{ik},y_{jk}')) =  \prod_{j}\EEE_{y'}\exp(\lambda\sum_{i}a_{ijk}h_2(y_{ik},y_{jk}')) \\
    \lesssim& \prod_{j}\exp\left( \frac{\lambda^2L_k\sum_{i}a_{ijk}^2}{2(1-c\lambda L_{k})} \right)
    \leq \exp\left( \frac{\lambda^2 L_{\max}\|\A_k\|_F^2}{2(1-c\lambda L_{\max})} \right) \lesssim \exp\left( \frac{\lambda^2 L_{\max}|B(\ppi,\ppi^*)\cap \PPP_k|}{2(1-c\lambda L_{\max})} \right),
\end{aligned}
$$
where the first inequality is due to Bernstein-type bound (e.g. Proposition 2.10 in \cite{wainwright2019high}) along with the facts that 
$$
\max_{j}|\sum_{i}a_{ijk}h_2(y_{ik},y_{jk}')|\lesssim \|\A_{k}\|_{1} \lesssim L_{k}~~\text{and}~~ 
\var_{y'}(\sum_{i}a_{ijk}h_2(y_{ik},y_{jk}')) \leq L_k\sum_{i} a_{ijk}^2 \var_{y'}(h_2(y_{ik},y_{jk}')) \lesssim L_k\sum_{i}a_{ijk}^2.
$$
Then, for $0<\lambda<(2cL_{\max})^{-1}$, $\EEE_{y}\EEE_{y'}\exp(\lambda\sum_{i,j}a_{ijk}h_2(y_{ik},y_{jk}')) \lesssim \exp(\lambda^2 L_{\max}|B(\ppi,\ppi^*)\cap \PPP_k|)$,
which leads that $$
\begin{aligned}
\EEE\exp(\lambda \sum_{k=1}^pI_{k2}) =& \EEE\exp(\lambda\sum_{k=1}^p\sum_{i,j}a_{ijk}h_2(y_{ik},y_{jk})) \leq \prod_{k=1}^p\EEE_{y,y'}\exp(\lambda\sum_{i,j}a_{ijk}h_2(y_{ik},y_{jk}')) \\
\leq& \prod_{k=1}^p\exp(\lambda^2 L_{\max}|B(\ppi,\ppi^*)\cap \PPP_k|) = \exp(\lambda^2 L_{\max}\sum_{k=1}^p|B(\ppi,\ppi^*)\cap \PPP_k|).
\end{aligned}
$$
Therefore, for $0<\lambda<(2cL_{\max})^{-1}$, $$
\begin{aligned}
\Pr(\sum_{k=1}^pI_{k2}>\epsilon) \leq& e^{-\lambda \epsilon}\EEE\exp(\lambda \sum_{k=1}^pI_{k2}) \leq \exp(-\lambda\epsilon + \lambda^2 L_{\max}\sum_{k=1}^p|B(\ppi,\ppi^*)\cap \PPP_k|)\\
    \lesssim& 
    \left\{
    \begin{aligned}
        &\exp\left( \frac{-c\epsilon^2}{L_{\max}\sum_{k=1}^p|B(\ppi,\ppi^*)\cap\PPP_k|} \right),\quad&&\text{if}\quad\frac{\epsilon}{2L_{\max}\sum_{k=1}^p|B(\ppi,\ppi^*)\cap\PPP_k|} < \frac{1}{2cL_{\max}} \\
        &\exp\left(\frac{-\epsilon}{2cL_{\max}}\right),  \quad&&\text{otherwise},
    \end{aligned}
    \right. \\
\lesssim& \exp\left( \frac{-c\epsilon^2}{L_{\max}\sum_{k=1}^p|B(\ppi,\ppi^*)\cap \PPP_k| + L_{\max}\epsilon} \right),
\end{aligned}
$$ and similar result hold for $\Pr(\sum_{k=1}^pI_{k2}<-\epsilon)$. Then, 
\begin{equation}\label{eq:2nd order}
\begin{aligned}
    \Pr\left(|\sum_{k=1}^pI_{k2}|>\epsilon\right) \lesssim& \exp\left( \frac{-c\epsilon^2}{L_{\max}\sum_{k=1}^p|B(\ppi,\ppi^*)\cap \PPP_k| + L_{\max}\epsilon} \right)
\end{aligned}
\end{equation}
Recall that $G(\ppi) = \sum_{k=1}^p I_k = \sum_{k=1}^p I_{k_1} + \sum_{k=1}^p I_{k_2}$,
combining \eqref{eq:1st order} and \eqref{eq:2nd order}, we conclude that 
$$
\Pr\left(\sup_{\ppi:\sum_{k=1}^p|B(\ppi,\ppi^*)\cap\PPP_k|=t}|G(\ppi)|>\epsilon\right) \lesssim n!\exp\left( \frac{-\epsilon^2}{L_{\max}t + L_{\max}\epsilon} \right)  \lesssim \exp\left(n\log n - \frac{\epsilon^2}{L_{\max}t + L_{\max}\epsilon} \right),
$$
which implies that as $p\to\infty$,
$$
\sup_{\ppi:\sum_{k=1}^p|B(\ppi,\ppi^*)\cap\PPP_k|=t}|G(\ppi)| =O_p\Big(\max\big\{\sqrt{tL_{\max}n\log n}, L_{\max}n\log n\big\}\Big).
$$
For any $\ppi$ and $t$ such that $\sum_{k=1}^p|B(\ppi,\ppi^*)\cap\PPP_k|=t$ and 
\begin{equation}\label{eq:condition}
\frac{(\sum_{k=1}^p\sum_{(i,j)\in B(\ppi,\ppi^*)\cap \PPP_k}q_{ijk})^2}{\sum_{k=1}^p|B(\ppi,\ppi^*)\cap\PPP_k|} \gg L_{\max}n\log n,
\end{equation}
then according to \eqref{eq:signal}, we obtain that 
$$
\EEE L(\ppi^*) - \EEE L(\ppi) = 2\sum_{k=1}^p\sum_{(i,j)\in B(\ppi^*,\ppi)\cap\PPP_k} q_{ijk}\gg \sqrt{tL_{\max}n\log n} + L_{\max}n\log n,
$$
which implies that, with probability approaching 1 as $p\to\infty$,
$$
L(\ppi^*) - L(\ppi) > \EEE L(\ppi^*) - \EEE L(\ppi) - \sup_{\ppi:\sum_{k=1}^p|B(\ppi,\ppi^*)\cap\PPP_k|=t}|G(\ppi)| > 0.
$$
To conclude, for any $\ppi\in S_n$ such that \eqref{eq:condition} holds, we have $L(\ppi^*) > L(\ppi)$ with probability approaching 1. Since $L(\widehat\ppi) \geq L(\ppi^*)$, then as $p\to\infty$,
$$
\frac{(\sum_{k=1}^p\sum_{(i,j)\in B(\widehat\ppi,\ppi^*)\cap \PPP_k}q_{ijk})^2}{\sum_{k=1}^p|B(\widehat\ppi,\ppi^*)\cap\PPP_k|} = O_p(L_{\max}n\log n).
$$
\end{proof}





\begin{proof}[Proof of Proposition~\ref{prop:KT}]
    Under the first condition, according to Proposition~\ref{prop:connect} and Theorem~\ref{thm:upper}, $$
    \tau(\widehat\ppi,\ppi^*) = O_p\left(\frac{L_{\max}n\log n}{\bar Nq_{\min}^2}\right) = o_p(n^2).
    $$


Under the second condition, note that $\widehat\ppi = \rank(\yy)$, where $\yy = (y_{i1},...,y_{in})$. For any $\xx\in\R^n$, define 
$$
H(\xx) = \frac{\tau(\rank(\xx),\ppi^*)}{n(n-1)/2} = \frac{\sum_{i,j\in[n]}1_{\{x_i>x_j\}}1_{\{\pi_i^*<\pi_j^*\}}}{n(n-1)/2}.
$$ 
Note that $H(\yy) = \tau(\widehat\ppi,\ppi^*)/\binom{n}{2}$ and $\EEE H(\yy) = \sum_{i,j\in[n]}\Pr(y_{i1} > y_{j1})1_{\{\pi_i^*<\pi_j^*\}}/\binom{n}{2} = o(1)$.
For any $\xx, \xx'\in\R^n$ which only differ in the $i$-th position, we have 
$$
|H(\xx) - H(\xx')| \leq \frac{n-1}{n(n-1)/2} = \frac{2}{n},
$$
which means $H(\cdot)$ satisfies the bounded differences property with all bounds being $2/n$. By McDiarmid's inequality, we have 
$$
\Pr(|H(\yy) - \EEE H(\yy)| > \epsilon) \leq 2\exp\left(-\frac{2\epsilon^2}{\sum_{i=1}^n\frac{4}{n^2}}\right) =2\exp(-n\epsilon^2/2),
$$
which implies $ \tau(\widehat\ppi,\ppi^*)/\binom{n}{2} = H(\yy) \leq \EEE H(\yy) + o_p(n^{-1/2}) = o_p(1)$. 
\end{proof}

\begin{proof}[Proof of Proposition~\ref{prop:cond}]
    The proof stands by the fact that we only rely on  $L(\widehat\ppi)>L(\ppi^*)$ in the proof of Theorem~\ref{thm:upper}.
\end{proof}

\begin{proof}[Proof of Theorem~\ref{thm:lower}]
    Given $k\in[p]$ and $\ppi\in S_n^+$, define $\ppi_{\CCC_k}$ as the relative ranking of $\CCC_k$ based on $\ppi$. For example, if $\CCC_k = \{2,3,4\}$ and $\ppi = (3,5,2,4,...)$, then $\ppi_{\CCC_k} = (3,1,2)$. 
    For $i\in[n]$, define $$
    h_i(y) = \frac{1}{2} + \frac{n}{2}1_{\{y\in[\frac{i-1}{n},\frac{i}{n})\}}
    $$ to be the density function for the mixture distribution $0.5\text{Unif}(0,1)+0.5\text{Unif}\big(\frac{i-1}{n},\frac{i}{n}\big)$.
    For any vector $\s = (i_1,...,i_L)\in[n]^L$ with different entries, define a joint density function $h_{\s}(y_1,...,y_L) = h_{i_1}(y_1)\cdots h_{i_L}(y_L)$.
    
    We further define a parameter space for the joint distribution of $\DDDD$ as 
    $$
        \widetilde\GGG = \Bigg\{G\in\GGG: \text{under }G, \ppi_{\obs,k}\mid \CCC_k\overset{d}{=}\rank(\yy_{\CCC_k,k}), \text{ with } \yy_{\CCC_k,k}\sim h_{\ppi(G)_{\CCC_k}}(\cdot)\Bigg\}
    $$
    Note that there is a bijection between $\widetilde \G$ and $S_n$. For any $\ppi\in S_n$, we use $G_{\ppi}$ to denote the corresponding $G\in\widetilde\GGG$ such that $\ppi(G_{\ppi}) = \ppi$.
    Since the raters are independent, we have 
    $$
    G_{\ppi} = G_{\ppi,1}\cdots G_{\ppi,p}
    $$ 
    with $G_{\ppi,k}$ being the distribution of $(\CCC_k,\ppi_{\obs,k})$.

    Under $G_{\ppi}$, note that $\Pr(y_{ik}< y_{jk}) = 3/4$ if $\ppi_i<\ppi_j$, and thus 
    \begin{equation}\label{eq:prob gap}
    q_{ijk} = \frac{1}{2} \quad\text{for all}~i<j\in[n]~ \text{and}~ k\in[p].
    \end{equation}

We first control the KL divergence between \(G_{\ppi}\) and \(G_{\ppi'}\). Let
\(\nu_k\) denote the sampling distribution of \(\CCC_k\), which does not depend
on \(\ppi\). For a fixed subset \(\CCC\in S_{n,k}\), let
\(P_{\ppi,\CCC}\) be the distribution of \(\rank(y_{\CCC,k})\) when
\(y_{\CCC,k}\sim h_{\ppi_\CCC}\). Then
\[
    G_{\ppi,k}(\CCC,r)=\nu_k(\CCC)P_{\ppi,\CCC}(r).
\]
Since \(\nu_k\) is the same under \(G_{\ppi}\) and \(G_{\ppi'}\), we have
\[
\begin{aligned}
    D_{\KL}(G_{\ppi}\|G_{\ppi'})
    &=
    \sum_{k=1}^p D_{\KL}(G_{\ppi,k}\|G_{\ppi',k})                                      \\
    &=
    \sum_{k=1}^p\sum_{\CCC\in S_{n,k}}\nu_k(\CCC)
    D_{\KL}(P_{\ppi,\CCC}\|P_{\ppi',\CCC}).
\end{aligned}
\]
By the data processing inequality (Theorem 2.8.1 in \cite{cover1999elements}),
\[
    D_{\KL}(P_{\ppi,\CCC}\|P_{\ppi',\CCC})
    \le
    D_{\KL}(h_{\ppi_\CCC}\|h_{\ppi'_\CCC}).
\]
A direct calculation gives
\[
    D_{\KL}(h_a\|h_b)=\frac{1}{2}\log(n+1),
    \qquad a\neq b.
\]
Consequently,
\[
\begin{aligned}
    D_{\KL}(P_{\ppi,\CCC}\|P_{\ppi',\CCC})
    &\le
    \frac{1}{2}\log(n+1)
    \left|\{i\in[L_k]:\pi_{\CCC,i}\neq \pi'_{\CCC,i}\}\right|       \\
    &\le
    \tau(\ppi_\CCC,\ppi'_\CCC)\log(n+1),
\end{aligned}
\]
where the second inequality is due to the fact that $\left|\{i\in[L_k]:\pi_{\CCC,i}\neq \pi'_{\CCC,i}\}\right| \leq 2\tau(\ppi_{\CCC},\ppi'_{\CCC})$, since every $i\in[L_k]$ such that $\pi_{\CCC,i}\neq \pi'_{\CCC,i}$, there must exist a pair in $B(\ppi_{\CCC},\ppi'_{\CCC})$ with one coordinate being $i$, and each pair has two coordinates.

It follows that
\[
\begin{aligned}
    D_{\KL}(G_{\ppi}\|G_{\ppi'})
    &\le
    \log(n+1)
    \sum_{k=1}^p
    \sum_{\CCC\in S_{n,k}}\nu_k(\CCC)\tau(\ppi_\CCC,\ppi'_\CCC)        \\
    &=
    \log(n+1)
    \sum_{k=1}^p
    \sum_{(i,j)\in B(\ppi,\ppi')}
    \Pr(\{i,j\}\subseteq \CCC_k).
\end{aligned}
\]
By Assumption~\ref{ass:non_uniform},
\[
    \Pr(\{i,j\}\subseteq \CCC_k)
    \le
    \delta\frac{L_k(L_k-1)}{n(n-1)}.
\]
Therefore,
\begin{equation}\label{eq:KL1}
    D_{\KL}(G_{\ppi} \| G_{\ppi'})
    \le
    \delta \tau(\ppi,\ppi')
    \frac{\sum_{k=1}^p L_k(L_k-1)}{n(n-1)}
    \log(n+1).
\end{equation}
On the other hand, using the same data processing argument directly gives
\begin{equation}\label{eq:KL2}
\begin{aligned}
    D_{\KL}(G_{\ppi}\|G_{\ppi'})
    &\le
    \sum_{k=1}^p
    \sum_{\CCC\in S_{n,k}}\nu_k(\CCC)
    D_{\KL}(h_{\ppi_\CCC}\|h_{\ppi'_\CCC})        \\
    &\le
    \sum_{k=1}^p \frac{1}{2}L_k\log(n+1)\sum_{\CCC\in S_{n,k}}\nu_k(\CCC) = \frac{1}{2}L_k\log(n+1).
\end{aligned}
\end{equation}
Combining \eqref{eq:KL1} and \eqref{eq:KL2}, we have
\begin{equation}\label{eq:KL upper}
    D_{\KL}(G_{\ppi}\|G_{\ppi'})
    \le
    \min\left\{
    \delta \tau(\ppi,\ppi')
    \frac{\sum_{k=1}^p L_k(L_k-1)}{n(n-1)}
    \log(n+1),~
    \frac{1}{2}\sum_{k=1}^p L_k\log(n+1)
    \right\}.
\end{equation}

We next relate the modified Kendall's tau loss to the standard Kendall's tau
loss under this construction. By \eqref{eq:prob gap}, for any two rankings \(\ppi,\ppi'\),
\[
    \tau_m(\ppi,\ppi')
    =
    \frac{1}{4}
    \sum_{k=1}^p
    |B(\ppi,\ppi')\cap \PPP_k|.
\]
Taking expectation over the random design and using the lower bound in
Assumption~\ref{ass:non_uniform} yields
\[
\begin{aligned}
    \EEE_G\tau_m(\ppi,\ppi')
    &=
    \frac{1}{4}
    \sum_{k=1}^p
    \sum_{(i,j)\in B(\ppi,\ppi')}
    \Pr(\{i,j\}\subseteq C_k)                      \\
    &\ge
    \frac{1}{4\delta}
    \frac{\sum_{k=1}^p L_k(L_k-1)}{n(n-1)}
    \tau(\ppi,\ppi').
\end{aligned}
\]
Thus,
\begin{equation*}
\begin{aligned}
    \min_{\check\ppi=T(D)}
    \max_{G\in\mathcal G}
    \EEE_G\tau_m(\check\ppi,\ppi(G))
    &\ge
    \min_{\check\ppi=T(D)}
    \max_{G\in\widetilde{\mathcal G}}
    \EEE_G\tau_m(\check\ppi,\ppi(G))            \\
    &\ge
    \frac{\sum_{k=1}^p L_k(L_k-1)}
    {4\delta n(n-1)}
    \min_{\check\ppi=T(D)}
    \max_{G\in\widetilde{\mathcal G}}
    \EEE_G\tau(\check\ppi,\ppi(G)).
\end{aligned}
\end{equation*}

Let \(\ppi_0=(1,2,\ldots,n)\in S_n\). For any
\(r=1,\ldots,{n\choose 2}\), define
\[
    B_{\mathrm{KT}}(r)=\{\ppi\in S_n:\tau(\ppi,\ppi_0)\le r\}.
\]
Then
\begin{equation}\label{eq:modified lower}
\begin{aligned}
    \min_{\check\ppi=T(D)}
    \max_{G\in\mathcal G}
    \EEE_G\tau_m(\check\ppi,\ppi(G)) \ge
    \frac{\sum_{k=1}^p L_k(L_k-1)}
    {4\delta n(n-1)}
    \min_{\check\ppi=T(D)}
    \max_{\substack{G\in\widetilde{\mathcal G}\\
    \ppi(G)\in B_{\mathrm{KT}}(r)}}
    \EEE_G\tau(\check\ppi,\ppi(G)).
\end{aligned}
\end{equation}

We require the following lemma, the first part of which is basically the Proposition 3 in \cite{mao2018minimax}, with minor technical differences.

\begin{lemma}\label{lem:packing}
    For $r\in[\binom{n}{2}]$ and $\epsilon = r/100$, 
    $$
    \begin{aligned}
        &\log M(\epsilon,B_{\KT}(r)) \geq n,\quad&&\text{if}~r\ge 10n \\
        &\log M(\epsilon,B_{\KT}(r))\geq r, \quad&&\text{if}~r\leq n/10.
    \end{aligned}
    $$
\end{lemma}

By Fano's lemma and \eqref{eq:KL upper},
\begin{equation}\label{eq:fano}
\begin{aligned}
&\min_{\check\ppi=T(D)}
\max_{\substack{G\in\widetilde{\mathcal G}\\
\ppi(G)\in B_{\mathrm{KT}}(r)}}
\EEE_G\tau(\check\ppi,\ppi(G))                                      \\
&\qquad \ge
\epsilon
\left(
1-
\frac{
\max_{\ppi,\ppi'\in B_{\mathrm{KT}}(r)}
D_{\KL}(G_{\ppi}\|G_{\ppi'})
+\log 2
}{
\log M(\epsilon,B_{\mathrm{KT}}(r))
}
\right)                                                               \\
&\qquad \ge
\frac{r}{100}
\left(
1-
\frac{
\min\left\{
2\delta r
\frac{\sum_{k=1}^p L_k(L_k-1)}{n(n-1)}
\log(n+1),
\;
\sum_{k=1}^p L_k\log(n+1)
\right\}
+\log 2
}{
\log M(\epsilon,B_{\mathrm{KT}}(r))
}
\right),
\end{aligned}
\end{equation}
where we used \(\tau(\ppi,\ppi')\le 2r\) for
\(\ppi,\ppi'\in B_{\mathrm{KT}}(r)\).

To get the minimax lower bound for $\EEE_G\tau(\check\ppi,\ppi(G))$, we need to find the maximal $r$ such that 
$$
\min\Big\{\frac{2\delta r\sum_{k=1}^pL_k(L_k-1)\log(n+1)}{n(n-1)},  \frac{1}{2}\sum_{k=1}^p L_k \log (n+1) \Big\} \leq \frac{1}{2}\log M(\epsilon,B_{KT}(r)).
$$

\textbf{Case I:} $\frac{2\delta r\sum_{k=1}^pL_k(L_k-1)\log(n+1)}{n(n-1)} \leq \frac{1}{2}\log M(\epsilon, B_{KT}(r))$.
For $r\geq 10n$, a sufficient condition is
$$
r\leq \frac{n^2(n-1)}{4\delta\sum_{k=1}^pL_k(L_k-1)\log(n+1)}.
$$
Therefore, 
\begin{equation}\label{eq:minimax11}
\min\limits_{\check\ppi = T(\DDDD) }\max\limits_{\substack{G\in\widetilde\GGG \\\ppi(G)\in B_{KT}(r)}}\EEE_G\tau(\check\ppi,\ppi(G))\gtrsim \frac{n^3}{\delta\sum_{k=1}^pL_k^2\log n},\quad\text{if}~~ \sum_kL_k(L_k-1)\leq \frac{n(n-1)}{40\delta\log(n+1)}.
\end{equation}
For $r\leq n/10$, a sufficient condition is
$$
\frac{\delta\sum_{k}L_k(L_k-1)\log(n+1)}{n(n-1)}\leq \frac{1}{4}.
$$
Therefore, 
\begin{equation}\label{eq:minimax12}
 \min\limits_{\check\ppi = T(\DDDD) }\max\limits_{\substack{G\in\widetilde\GGG \\\ppi(G)\in B_{KT}(r)}}\EEE_G\tau(\check\ppi,\ppi(G))\gtrsim n,\quad\text{if}~~ \sum_{k}L_k(L_k-1)\leq \frac{n(n-1)}{4\delta\log(n+1)}.   
\end{equation}
Note that \eqref{eq:minimax11} is essentially tighter than \eqref{eq:minimax12}.

\textbf{Case II: $\frac{1}{2}\sum_k L_k\log (n+1)\leq \frac{1}{2}\log M(\epsilon, B_{KT}(r)).$} 
For $r\geq 10n$, a sufficient condition is $\sum_k L_k \log (n+1) \leq n$.
Therefore, 
\begin{equation}\label{eq:minimax21}
\min\limits_{\check\ppi = T(\DDDD) }\max\limits_{\substack{G\in\widetilde\GGG \\\ppi(G)\in B_{KT}(r)}}\tau(\check\ppi,\ppi(G))\asymp n^2,\quad\text{if}~~ 
\sum_k L_k \leq \frac{n}{\log (n+1)}.
\end{equation}
For $r\leq n/10$, a sufficient condition is $\sum_{k}L_k\log (n+1)\leq r$.
Therefore, 
\begin{equation}\label{eq:minimax22}
\min\limits_{\check\ppi = T(\DDDD) }\max\limits_{\substack{G\in\widetilde\GGG \\\ppi(G)\in B_{KT}(r)}}\tau(\check\ppi,\ppi(G))\gtrsim n,\quad\text{if}~~ \sum_kL_k\leq \frac{n}{10\log (n+1)}.
\end{equation}
Note that \eqref{eq:minimax21} is essentially tighter than \eqref{eq:minimax22}.

According to \eqref{eq:modified lower}, \eqref{eq:minimax11} and \eqref{eq:minimax21}, we obtain 
$$
\min_{\check\ppi = T(\DDDD) }\max_{G\in\GGG}\tau_{m}(\check\ppi,\ppi) \gtrsim \left\{
    \begin{aligned}
        &\frac{n}{\delta^2\log n},\quad&&\text{if}\quad \sum_k L_k(L_k-1) \leq \frac{n(n-1)}{40\delta\log(n+1)},\\
        &\frac{\sum_{k}L_k^2}{\delta},\quad&&\text{if}\quad \sum_{k}L_k\leq \frac{n}{\log(n+1)}.
    \end{aligned}
\right.
$$
\end{proof}

\begin{proof}[Proof of Lemma~\ref{lem:tech}]
Since $a_{ijk} = -a_{jik}$ and $z_{ijk} = 1_{\{y_{ik} < y_{jk}\}}-1_{\{y_{ik}> y_{jk}\}} = -z_{jik}$, then
    $$
    \begin{aligned}
        I_k =& \sum_{j:i<j} a_{ijk}(z_{ijk}-\EEE z_{ijk}) + \sum_{l:l<i}a_{lik}(z_{lik}-\EEE z_{lik})\\ 
        =& \sum_{j:i<j} a_{ijk}(z_{ijk}-\EEE z_{ijk}) + \sum_{l:l<i}a_{ilk}(z_{ilk}-\EEE z_{ilk}) = \sum_{j:j\neq i} a_{ijk}(z_{ijk}-\EEE z_{ijk}),
    \end{aligned}
    $$ which leads to the first equation immediately by taking conditional expectation. The second equation is due to that $\EEE (z_{ijk}\mid y_{ik}) = -\EEE(z_{jik}\mid y_{ik})$ and thus $a_{ijk}\EEE(z_{ijk}\mid y_{ik}) = a_{jik}\EEE(z_{jik}\mid y_{ik})$. The third equation is due to that $$
        \sum_{i<j}a_{ijk}(z_{ijk}-\EEE z_{ijk}) = \sum_{i<j}a_{jik}(z_{jik}-\EEE z_{jik})\quad \text{and}\quad \sum_{i<j}a_{ijk}(z_{ijk}-\EEE z_{ijk}) + \sum_{i>j}a_{ijk}(z_{ijk}-\EEE z_{ijk}) = \sum_{i\neq j }a_{ijk}(z_{ijk}-\EEE z_{ijk}).
    $$
\end{proof}

\begin{proof}[Proof of Lemma~\ref{lem:packing}]
 According to Proposition 3 in \cite{mao2018minimax}, if $r\geq 10n$, then
 $$
 \begin{aligned}
    \log M(\epsilon, B_{KT}(r))\geq& n\log(\frac{r}{n+\epsilon}) - 2n = n\log(100 - \frac{100n}{n+\frac{r}{100}}) - 2n\\ 
    \geq& n\log(100 - \frac{100n}{n+\frac{n}{10}}) -2n >n\log 50 - 2n > n.
\end{aligned}
 $$ 
For the second result, we first claim that 
\begin{equation}\label{eq:entropy}
\begin{aligned}
    r\log \frac{n-1}{r} \leq \log|B_{KT}(r)| \leq& r\log \frac{n+r-1}{r}+r\quad&&\text{for}\quad 1\leq r < n.
\end{aligned}
\end{equation}
If this holds true and $r\leq n/10$, then $(n-1)/r\geq 9$.
Let $N(\epsilon,B_{KT}(r))$ be the $\epsilon$-covering number for $B_{KT}(r)$.
Then 
$$
\begin{aligned}
\log M(\epsilon, B_{KT}(r))\geq& \log N(\epsilon,B_{KT}(r)) \geq \log|B_{KT}(r)| - \log|B_{KT}(\epsilon)| \\
\geq & r\log\frac{n-1}{r} - \epsilon\log\frac{n+\epsilon-1}{\epsilon} -\epsilon =r\log\frac{n-1}{r} - \frac{r}{100}\log(\frac{100(n-1)}{r}+1) - \frac{r}{100} \\
=&r(\log9-\frac{\log901}{100}-\frac{1}{100})>r.
\end{aligned}
$$

It remains to prove \eqref{eq:entropy}. 
We represent each permutation $\ppi \in S_n$ by its inversion vector $\aaaa = (a_1, a_2, \dots, a_n)$ \cite{zhou2014systematic}, defined by $a_i = |\{j>i:\pi_j<\pi_i\}|$ for $i\in[n]$.
The total number of inversions is given by $\tau(\ppi, \ppi_0) = \sum_{i=1}^n a_i$, with the constraints $0 \le a_i \le n - i$ for each $i \in [n]$. 
To establish the upper bound in \eqref{eq:entropy}, note that the number of permutations with at most $r$ inversions is bounded above by the number of non-negative integer solutions to: 
$$
\sum_{i=1}^{n-1} a_i \le r, \quad a_i \ge 0,
$$ which is $\binom{(n-1) + r}{r}$.
Then, 
$$
\log |B_{KT}(r)| = \log\binom{n+r-1}{r}\leq \log\left( \frac{(n+r-1)e}{r} \right)^r = r \log \frac{n+r-1}{r} + r,
$$
where the third inequality is due to that $\binom{N}{k} \le \left( \frac{Ne}{k} \right)^k$.

We then turn to establish the lower bound in \eqref{eq:entropy} by considering the set of permutations $\mathcal{S} \subseteq B_{KT}(r)$ whose inversion vectors $\aaaa = (a_1, \dots, a_n) \subseteq\{0, 1\}^n$. 
Specifically, a permutation $\pi$ belongs to $\mathcal S$ if its inversion vector satisfies:
\begin{enumerate}
    \item $a_i \in \{0, 1\}$ for all $i \in [n-1]$.
    \item $\sum_{i=1}^{n-1} a_i = r$.
\end{enumerate}
The number of ways to construct a binary inversion vector with exactly $r$ ones is the number of ways to choose $r$ indices from the available $n-1$ positions: $|B_{KT}(r)| \ge |\mathcal{S}| = \binom{n-1}{r}$. Then 
$$
\log |B_{KT}(r)| \ge\log \binom{n-1}{r}\geq \log \left(\frac{n-1}{r}\right)^r = r\log\left(\frac{n-1}{r}\right), 
$$ 
where the second inequality is due to that $\binom{N}{k} \ge \left( \frac{N}{k} \right)^k$.
\end{proof}

\bigskip\bigskip

\bibliographystyle{plainnat} 
\bibliography{reference}
\end{document}